\documentclass[%
 reprint,
groupedaddress,
 amsmath,amssymb,
 aps,
prx,
]{revtex4-2}

\usepackage{graphicx}
\usepackage{dcolumn}
\usepackage{bm}
\usepackage{hyperref}

\usepackage{lipsum}
\usepackage{physics}
\usepackage{tikz}
\usetikzlibrary{decorations.pathmorphing}
\usepackage{bbm}
\usepackage{mathrsfs}
\usepackage{siunitx}
\def \d{\partial}
\def \Ztwo{$\mathbb{Z}_2$~}
\def \hilb{\mathscr{H}}
\def \P{\langle \hat{P} \rangle}
\def \Q{\langle \hat{Q} \rangle}
\def \PFM{\langle \hat{P}_\text{FM} \rangle}
\def \QFM{\langle \hat{Q}_\text{FM} \rangle}

\definecolor{mymagenta}{HTML}{AD4096}
\definecolor{myblue}{HTML}{313180}
\newcommand{\scale}{0.75}
\newcommand{\lw}{0.625pt}

\def \dimer{
    \begin{tikzpicture}[baseline=-0.6ex, line width=\lw]
        \draw (0,0) -- (0.3,0);
        \draw[draw=red, draw opacity=1, fill=red, fill opacity=0.15] (0.15,0) ellipse (0.15 and 0.05);
    \end{tikzpicture}
}

\def \monomer{
    \begin{tikzpicture}[baseline=-0.6ex, line width=\lw]
        \draw (0,0) -- (0.3,0);
    \end{tikzpicture}
}

\newlength{\tablevspace}
\newlength{\textvspace}
\newlength{\y}

\def \ggg {
    \ifmmode
        \setlength{\y}{\textvspace}
    \else
        \setlength{\y}{\tablevspace}
    \fi
    \begin{tikzpicture}[baseline=\y, line width=\lw, scale=\scale]
        \draw (0,0) -- (0.5,0) -- (0.25,{sqrt(3)/4}) -- (0,0);
    \end{tikzpicture}
}

\def \ggr {
    \ifmmode
        \setlength{\y}{\textvspace}
    \else
        \setlength{\y}{\tablevspace}
    \fi
    \begin{tikzpicture}[baseline=\y, line width=\lw, scale=\scale]
        \draw (0,0) -- (0.5,0) -- (0.25,{sqrt(3)/4}) -- (0,0);
        \draw[draw=red, draw opacity=1, fill=red, fill opacity=0.15, cm={cos(60) ,-sin(60) ,sin(60) ,cos(60) ,(0.25,{sqrt(3)/4})}] (0.25,0) ellipse (0.25 and 0.05);
    \end{tikzpicture}
}

\def \grg {
    \ifmmode
        \setlength{\y}{\textvspace}
    \else
        \setlength{\y}{\tablevspace}
    \fi
    \begin{tikzpicture}[baseline=\y, line width=\lw, scale=\scale]
        \draw (0,0) -- (0.5,0) -- (0.25,{sqrt(3)/4}) -- (0,0);
        \draw[draw=red, draw opacity=1, fill=red, fill opacity=0.15, cm={cos(60) ,sin(60) ,-sin(60) ,cos(60) ,(0,0)}] (0.25,0) ellipse (0.25 and 0.05);
    \end{tikzpicture}
}

\def \rgg {
    \ifmmode
        \setlength{\y}{\textvspace}
    \else
        \setlength{\y}{\tablevspace}
    \fi
    \begin{tikzpicture}[baseline=\y, line width=\lw, scale=\scale]
        \draw (0,0) -- (0.5,0) -- (0.25,{sqrt(3)/4}) -- (0,0);
        \draw[draw=red, draw opacity=1, fill=red, fill opacity=0.15] (0.25,0) ellipse (0.25 and 0.05);
    \end{tikzpicture}
}

\def \PDrawing {
\begin{tikzpicture}[baseline=\textvspace, line width=\lw, scale=\scale]
    \draw (0,0) -- (0.5,0) -- (0.25,{sqrt(3)/4}) -- (0,0);
    \draw[mymagenta,thick,dashed, dash pattern={on 2pt off 1pt},cm={1,0,0,1, (0,sqrt(3/128))}] (0,0) -- (0.5,0);
\end{tikzpicture}
}

\def \QDrawing {
\begin{tikzpicture}[baseline=\textvspace, line width=\lw, scale=\scale]
    \draw (0,0) -- (0.5,0) -- (0.25,{sqrt(3)/4}) -- (0,0);
    \draw[myblue,very thick,decorate,decoration={snake,amplitude=1.5,segment length=3}] (0,0) -- (0.5,0);
\end{tikzpicture}
}

\newcommand{\transstar}[2]
{
    
    \coordinate  (A) at ({0.25*0.5 *1/2+#1*0.5},{-0.25 *sqrt(3)/2*0.5+#2*0.5});
    \coordinate  (B) at ({ 0.25 *sqrt( 6.25 )*0.5+0.25*0.5+#1*0.5},{ 0.25 *sqrt( 18.75 )*0.5+#2*0.5});
    \coordinate  (C) at ({-0.25 *sqrt( 2.25 )*0.5-0.25*0.5+#1*0.5},{ 0.25 *sqrt( 18.75 )*0.5+#2*0.5});
    \draw (A) -- (B) -- (C) -- (A);
    
    \coordinate  (A) at ({ 0.25 *sqrt( 0.25 )*0.5+#1*0.5},{ 0.25 *sqrt( 18.75 )*0.5+ 0.25*0.5 *sqrt(3)+#2*0.5});
    \coordinate  (B) at ({ 0.25 *sqrt( 6.25 )*0.5+0.25*0.5+#1*0.5},{ 0.25 *sqrt( 0.75 )*0.5+#2*0.5});
    \coordinate  (C) at ({-0.25 *sqrt( 2.25 )*0.5-0.25*0.5+#1*0.5},{ 0.25 *sqrt( 0.75 )*0.5+#2*0.5});
    \draw (A) -- (B) -- (C) -- (A);
}

\def \PFMDrawing{
\langle \hat{P}_\text{FM}\rangle = \frac{\Bigl\langle
\begin{tikzpicture}[baseline=1.55ex, line width=\lw]
\transstar{0}{0}

\coordinate (A) at (0.25,0);
\coordinate (B) at (0.4375,{0.1875*sqrt(3)});
\coordinate (C) at ({0.25},{0.375*sqrt(3)});
\coordinate (D) at ({-0.125},{0});

\draw[mymagenta,thick,dashed, dash pattern={on 2pt off 1pt}, line width=1pt, rounded corners=0.25pt] (D) -- (A) -- (B) -- (C) ;

\end{tikzpicture} \Bigr\rangle }{ \sqrt{\Bigl|\Bigl\langle
\begin{tikzpicture}[baseline=1.55ex, line width=\lw]
\transstar{0}{0}

\coordinate (A) at (0.25,0);
\coordinate (B) at (0.4375,{0.1875*sqrt(3)});
\coordinate (C) at ({0.25},{0.375*sqrt(3)});
\coordinate (D) at ({-0.125},{0.375*sqrt(3)});
\coordinate (E) at ({-0.3125},{0.1875*sqrt(3)});
\coordinate (F) at ({-0.125},{0});

\draw[mymagenta,thick,dashed, dash pattern={on 2pt off 1pt}, line width=1pt, rounded corners=0.25pt] (A)--(B)--(C)--(D)--(E)--(F) -- cycle ;

\end{tikzpicture} \Bigr\rangle\Bigr|}}
}

\def \QFMDrawing{
\langle \hat{Q}_\text{FM}\rangle = \frac{\Bigl\langle
\begin{tikzpicture}[baseline=1.55ex, line width=\lw]
\transstar{0}{0}
\draw[myblue,thick,decorate,decoration={snake,amplitude=1,segment length=2}] ({ 0.25 *sqrt( 0.25 )*0.5-0.25*0.5 },{ 0.25 *sqrt( 0.75 )*0.5 })
-- ({ 0.25 *sqrt( 0.25 )*0.5+0.25*0.5 },{ 0.25 *sqrt( 0.75 )*0.5 })
-- ({ 0.25 *sqrt( 6.25 )*0.5 },{ 0.25 *sqrt( 0.75 )*0.5+sqrt(3)/4*0.5 })
-- ({ 0.25 *sqrt( 0.25 )*0.5+0.25*0.5 },{ 0.25 *sqrt( 18.75 )*0.5 });

\end{tikzpicture} \Bigr\rangle}{\sqrt{\Bigl|\Bigl\langle
\begin{tikzpicture}[baseline=1.55ex, line width=\lw]
\transstar{0}{0}
\draw[myblue,thick,decorate,decoration={snake,amplitude=1,segment length=2}] ({ 0.25 *sqrt( 0.25 )*0.5-0.25*0.5 },{ 0.25 *sqrt( 0.75 )*0.5 })
-- ({ 0.25 *sqrt( 0.25 )*0.5+0.25*0.5 },{ 0.25 *sqrt( 0.75 )*0.5 })
-- ({ 0.25 *sqrt( 6.25 )*0.5 },{ 0.25 *sqrt( 0.75 )*0.5+sqrt(3)/4*0.5 })
-- ({ 0.25 *sqrt( 0.25 )*0.5+0.25*0.5 },{ 0.25 *sqrt( 18.75 )*0.5 })
-- ({ 0.25 *sqrt( 0.25 )*0.5-0.25*0.5 },{ 0.25 *sqrt( 18.75 )*0.5 })
-- ({ -0.25 *sqrt( 2.25 )*0.5 },{ 0.25 *sqrt( 0.75 )*0.5+sqrt(3)/4*0.5 })
-- ({ 0.25 *sqrt( 0.25 )*0.5-0.25*0.5 },{ 0.25 *sqrt( 0.75 )*0.5 });

\end{tikzpicture} \Bigr\rangle\Bigr|}}
}

\def \firstnn{
\begin{tikzpicture}[baseline=0.65ex,rotate=90, line width=\lw, scale=\scale]
    \draw (0,0) -- (0.5,0) -- (0.25,{sqrt(3)/4}) -- (0,0);
    \draw[draw=red, draw opacity=0.5, fill=red, fill opacity=0.15, cm={cos(60) ,-sin(60) ,sin(60) ,cos(60) ,(0.25,{sqrt(3)/4})}] (0.25,0) ellipse (0.25 and 0.08);
    \draw[draw=red, draw opacity=0.5, fill=red, fill opacity=0.15, cm={cos(60) ,sin(60) ,-sin(60) ,cos(60) ,(0,0)}] (0.25,0) ellipse (0.25 and 0.08);
    \draw (0.25,{sqrt(3)/4}) -- (0,{sqrt(3)/2}) -- (0.5,{sqrt(3)/2}) -- (0.25,{sqrt(3)/4});
\end{tikzpicture}
}

\def \secondnn{
\begin{tikzpicture}[baseline=0.65ex,rotate=90, line width=\lw, scale=\scale]
    \draw (0,0) -- (0.5,0) -- (0.25,{sqrt(3)/4}) -- (0,0);
    \draw[draw=red, draw opacity=0.5, fill=red, fill opacity=0.15, cm={cos(60) ,sin(60) ,-sin(60) ,cos(60) ,(0,0)}] (0.25,0) ellipse (0.25 and 0.08);    
    \draw[draw=red, draw opacity=0.5, fill=red, fill opacity=0.15, cm={cos(60) ,-sin(60) ,sin(60) ,cos(60) ,(0,{sqrt(3)/2})}] (0.25,0) ellipse (0.25 and 0.08);
    \draw (0.25,{sqrt(3)/4}) -- (0,{sqrt(3)/2}) -- (0.5,{sqrt(3)/2}) -- (0.25,{sqrt(3)/4});
\end{tikzpicture}
}

\def \thirdnn{
\begin{tikzpicture}[baseline=0.65ex,rotate=90, line width=\lw, scale=\scale]
    \draw (0,0) -- (0.5,0) -- (0.25,{sqrt(3)/4}) -- (0,0);
    \draw[draw=red, draw opacity=0.5, fill=red, fill opacity=0.15, cm={cos(60) ,sin(60) ,-sin(60) ,cos(60) ,(0.25,{sqrt(3)/4})}] (0.25,0) ellipse (0.25 and 0.08);
    \draw[draw=red, draw opacity=0.5, fill=red, fill opacity=0.15, cm={cos(60) ,sin(60) ,-sin(60) ,cos(60) ,(0,0)}] (0.25,0) ellipse (0.25 and 0.08);
    \draw (0.25,{sqrt(3)/4}) -- (0,{sqrt(3)/2}) -- (0.5,{sqrt(3)/2}) -- (0.25,{sqrt(3)/4});
\end{tikzpicture}
}

\begin{document}


\title{Predicting Topological Entanglement Entropy in a Rydberg analog simulator}

\author{Linda Mauron}
\affiliation{
Institute of Physics, École Polytechnique Fédérale de Lausanne (EPFL), CH-1015 Lausanne, Switzerland
}
\affiliation{
Center for Quantum Science and Engineering, \'{E}cole Polytechnique F\'{e}d\'{e}rale de Lausanne (EPFL), CH-1015 Lausanne, Switzerland
}

\author{Zakari Denis}
\affiliation{
Institute of Physics, École Polytechnique Fédérale de Lausanne (EPFL), CH-1015 Lausanne, Switzerland
}
\affiliation{
Center for Quantum Science and Engineering, \'{E}cole Polytechnique F\'{e}d\'{e}rale de Lausanne (EPFL), CH-1015 Lausanne, Switzerland
}

\author{Jannes Nys}%
\affiliation{
Institute of Physics, École Polytechnique Fédérale de Lausanne (EPFL), CH-1015 Lausanne, Switzerland
}
\affiliation{
Center for Quantum Science and Engineering, \'{E}cole Polytechnique F\'{e}d\'{e}rale de Lausanne (EPFL), CH-1015 Lausanne, Switzerland
}

\author{Giuseppe Carleo}%
\email{giuseppe.carleo@epfl.ch}
\affiliation{
Institute of Physics, École Polytechnique Fédérale de Lausanne (EPFL), CH-1015 Lausanne, Switzerland
}
\affiliation{
Center for Quantum Science and Engineering, \'{E}cole Polytechnique F\'{e}d\'{e}rale de Lausanne (EPFL), CH-1015 Lausanne, Switzerland
}

\date{\today}

\begin{abstract}
    Predicting the dynamical properties of topological matter is a challenging task, not only in theoretical and experimental settings, but also numerically. This work proposes a variational approach based on a time-dependent correlated Ansatz, focusing on the dynamical preparation of a quantum-spin-liquid state on a Rydberg-atom simulator. 
    Within this framework, we are able to faithfully represent the state of the system throughout the entire dynamical preparation protocol. The flexibility of our approach does not only allow one to match the physically correct form of the Rydberg-atom Hamiltonian but also the relevant lattice topology. This is unlike previous numerical studies which were constrained to simplified versions of the problem through the modification of both the Hamiltonian and the lattice. Our approach further gives access to global quantities such as the topological entanglement entropy ($\gamma$), providing insight into the topological properties of the system. 
    This is achieved by the introduction of the time-dependent variational Monte Carlo (t-VMC) technique to the dynamics of topologically ordered phases. Upon employing a Jastrow variational Ansatz with a scalable number of parameters, we are able to efficiently extend our simulations to system sizes matching state-of-the-art experiments and beyond.
    Our results corroborate experimental observations, confirming the presence of topological order during the dynamical state-preparation protocol, and additionally deepen our understanding of topological entanglement dynamics.
    We show that, while the simulated state exhibits (global) topological order and local properties resembling those of a resonating-valence-bond (RVB) state, it lacks the latter's characteristic topological entanglement entropy signature $\gamma = \ln(2)$, irrespective of the degree of adiabaticity of the protocol.
\end{abstract}

\maketitle

\section{\label{sec:intro} Introduction}
In recent years, topological properties of matter have attracted increasing interest~\cite{chiu_classification_2016, anirban_15_2023, bansil_colloquium_2016, martin_topology_2019}. In condensed matter, this has stimulated extensive research on topological materials, all the more since the experimental observation of the quantum hall effect~\cite{tsui_twodimensional_1982, laughlin_anomalous_1983} and chiral spin states~\cite{kalmeyer_equivalence_1987,wen_chiral_1989}. Since then, the study of many other topological phenomena, such as quantum spin liquids~\cite{read_largen_1991}, attracted consistent interest. Beyond condensed matter, the possibility of leveraging the robustness to local perturbations of topologically ordered states was soon exploited in the context of quantum error correction. 
The introduction of the toric code model~\cite{kitaev_anyons_2006} served as a stepping stone for a new paradigm in quantum computation, known as \textit{topological quantum computing}~\cite{dennis_topological_2002, nayak_nonabelian_2008, kitaev_topological_2009, freedman_topological_2002}. The topological properties of their ground state, such as long-range entanglement, make them robust against local operations and thus suitable for fault-tolerant computations~\cite{castagnoli_notions_1993, shor_faulttolerant_1996, gottesman_theory_1998, kitaev_faulttolerant_2003, fowler_surface_2012}.

Among the various phases of matter exhibiting topological order in two dimensions, quantum spin liquids (QSLs) stand out as particularly intriguing. These are states in which the spins exhibit no magnetic order while presenting strong quantum correlations across the system and (anyonic) many-body excitations~\cite{anderson_resonating_1973, savary_quantum_2017, sachdev_quantum_2023}. A paradigmatic model for understanding this phenomenon is again the toric code mentioned earlier~\cite{kitaev_anyons_2006}, which is a lattice \Ztwo gauge theory~\cite{fradkin_phase_1979, wen_zoo_2017, sachdev_quantum_2023}. The ground state of this model is four-fold degenerate and its so-called \textit{vacuum} ground state corresponds to a resonating valence bond (RVB) state~\cite{anderson_resonating_1973}. 
Excitations of this vacuum ground state manifest themselves as the emergence of anyons $1,e,m,f$ on the lattice. Anyons are quasi-particles carrying a fractionalized quantum number~\cite{wilczek_quantum_1982, arovas_fractional_1984, nakamura_direct_2020} and therefore always appear in pairs. 
The particular properties of anyons make quantum spin liquids promising candidates for topologically protected quantum memory~\cite{wen_quantum_2007, fowler_surface_2012}. 

A key characteristic of topologically ordered systems is their entanglement entropy. In addition to the area-law scaling observed in typical gapped ground states~\cite{srednicki_entropy_1993, plenio_entropy_2005}, topological states exhibit a negative offset in their entanglement entropy, known as the \textit{topological entanglement entropy}, which serves as an unequivocal marker of their topological order~\cite{wen_topological_1995, witten_sitter_1998, hamma_bipartite_2005, kitaev_topological_2006, levin_detecting_2006}.
Characterizing topological states in physical systems remains a challenge both for experiments and numerical simulations. Very recently, several experimental platforms have aimed at preparing topologically ordered states, including neutral atoms~\cite{semeghini_probing_2021, kalinowski_nonabelian_2023, sun_engineering_2023}, trapped ions~\cite{iqbal_creation_2023} and superconducting platforms~\cite{song_demonstration_2018, king_observation_2018, satzinger_realizing_2021}. In particular, a dynamical preparation protocol for a quantum spin liquid was proposed and demonstrated in Rydberg atom experiments~\cite{semeghini_probing_2021}. However, probing topological states in experiments is a challenging endeavor.
Indeed, unambiguously validating the presence of topological order requires the use of large-scale probes, since local probes are inconclusive. In particular, the entanglement signature of topological order is hard to access experimentally, and the use of quantum state tomography~\cite{leonhardt_quantumstate_1995, banaszek_maximumlikelihood_1999, gross_quantum_2010, pogorelov_experimental_2017, lanyon_efficient_2017, shang_superfast_2017, titchener_scalable_2018, kyrillidis_provable_2018, palmieri_experimental_2020, wang_scalable_2020, rambach_robust_2021, gebhart_learning_2023, chen_when_2023}, replicas~\cite{islam_measuring_2015, elben_renyi_2018, elben_statistical_2019, brydges_probing_2019} and other techniques~\cite{eisert_quantitative_2007, li_measuring_2019, foletto_experimental_2020, kim_recovering_2023, vermersch_manybody_2023, lin_measuring_2024} is under ongoing study, aiming to generalize the analysis to larger systems and harder types of quantum states. These challenges make experiments often rely on classical numerical simulations to faithfully validate the presence of topological order in such systems.

From a numerical standpoint, dynamical simulations of quantum systems remain a significant challenge. In particular, standard approaches such as tensor networks~\cite{white_density_1992, white_densitymatrix_1993, cirac_renormalization_2009, daley_timedependent_2004} suffer from the growth of entanglement upon time evolution~\cite{vidal_efficient_2003, eisert_entanglement_2013, eisert_quantum_2015}. This growth necessitates increasingly demanding computational resources, restricting simulations to small-scale systems and specific two-dimensional geometries and boundary conditions, especially in the case of density matrix renormalization group (DMRG)~\cite{verstraete_density_2004, stoudenmire_studying_2012, evenbly_tensor_2011, schollwoeck_densitymatrix_2011, jiang_identifying_2012, haegeman_timedependent_2011}, namely narrow cylinders. 
Moreover, matching the setting of the physical platforms introduces major complexities, such as the long-range coupling in Rydberg-atom lattices~\cite{hastings_solving_2006, pirvu_matrix_2010, zaletel_timeevolving_2015}.

Thus, the development of accurate and efficient simulation techniques is of prime importance, and variational Monte Carlo constitutes an appealing alternative. 
Ranging from simple mean-field Ansätze to deep neural quantum states~\cite{carleo_solving_2017}, flexibility in the choice of the variational state allows unprecedented versatility in the structure of the wave function. Furthermore, this approach lifts the restriction on the degree of entanglement and provides the possibility for complete control of the geometry of the physical system, its boundaries, and its topology (e.g.\ genus). This is essential when numerically investigating topological states. 
Moreover, the time-dependent variational Monte Carlo (t-VMC) algorithm~\cite{carleo_localization_2012, schmitt_dynamical_2015, carleo_unitary_2017, schmitt_quantum_2020, ido_timedependent_2015, nys_abinitio_2024} provides with an efficient and scalable way of simulating unitary time evolution within a fixed variational family. In turn, this is instrumental in the ability to simulate the time-dependent protocols involved in the preparation of the experimental state.

In this work, we show that topologically ordered phases can be efficiently simulated through time-dependent variational Monte Carlo (t-VMC) techniques, while offering a great degree of control over the geometry of the physical system. Moreover, this approach gives direct access to the time-dependent wave function, allowing one to extract various quantities of interest, and, in particular, the topological entanglement entropy. We focus on the dynamical state preparation of a quantum spin liquid with a Rydberg-atom simulator. Our approach allows us to consider a realistic experimental set-up, from the exact long-range Hamiltonian intrinsic to Rydberg platforms, to the geometry and boundary conditions of the lattice. We show that a lightweight variational Ansatz with a number of parameters scaling only linearly with system size is able to faithfully represent both an ideal quantum spin liquid and the state of the system at all times of the state preparation. This allows us to scale our simulations beyond the lattice sizes reported in experiments, simulating quantum dynamics on lattices with up to $288$ atoms. Although our numerical results agree with those of the experimental realization, we can directly probe the dynamics of the topological entanglement, thereby unambiguously confirming the presence of topological order at the end of the protocol. We further analyze the dependence of the maximum value of this quantity on the adiabaticity of the dynamical quench, showing that such an approach should prove particularly relevant when designing and optimizing experimental protocols.

The structure of our work is as follows. A detailed overview of the theoretical framework on topological entanglement entropy is presented in Sec.~\ref{sec:TEE}. Afterward, the physical system under study and the numerical methods are presented in Sec.~\ref{sec:model} and Sec.~\ref{sec:methods} respectively. The latter includes a discussion of the accuracy of our numerical method on a small system where comparison to exact simulations is possible. 
In Sec.~\ref{sec:results} we discuss the topological operators during the time evolution, along with an in-depth analysis of the effect of noise and a comparison with experimental data.
Consequently, we prepare a logical state in Sec.~\ref{sec:logic} and verify some basic properties of a Pauli algebra. 
Finally, in Sec.~\ref{sec:entropy}, we analyze the topological order of dynamically prepared states by calculating their topological entanglement entropy under various dynamical protocols, finally leading to our conclusions in Sec.~\ref{sec:conclusion}.

\section{\label{sec:TEE} Topological Entanglement Entropy}

The entanglement between a region $A$ and its complement $\bar{A}$ may be quantified by the entropy $S(\rho_A)$ of its reduced density matrix $\rho_A = \text{Tr}_{\Bar{A}} \dyad{\psi}{\psi}$. This can be evaluated through the von Neumann entanglement entropy~\cite{vonneumann_mathematische_1996} $S(\rho_A) = -\mathrm{Tr}_A(\rho_A \ln\rho_A)$ but also with Rényi-$n$~\cite{renyi_measures_1961} entanglement entropies $S^{(n)}(\rho_A) = \frac{1}{1-n} \ln\left[\mathrm{Tr}_A( \rho_A^n )\right]$, which converge to the former for $n\to 1$~\cite{flammia_topological_2009}.
Although these quantities follow an area-law scaling for the ground state of any gapped Hamiltonian~\cite{eisert_colloquium_2010}, topologically ordered states present the following crucial correction to this generic behavior
\begin{equation}
    S(\rho_A) = \alpha L - \gamma + \mathcal{O}(1) \,\text{,}
\label{eq:area_law}
\end{equation}
where $L$ is the length of the boundary between the two regions $A$ and $\Bar{A}$, $\alpha$ is a non-universal factor, and the offset $-\gamma<0$ is the \textit{topological entanglement entropy} (TEE), a fundamental feature of such states. The last term vanishes in the thermodynamical limit.

The area-law behavior~\eqref{eq:area_law} serves as evidence of the local entanglement near the boundaries of the two regions, while the value of the topological entanglement entropy provides a direct indicator of long-range entanglement, and, consequently, topological order. The occurrence of a negative offset is a particularly intriguing characteristic of topological order.
In a gapped ground state scenario that does not present topological order, a local unitary suffices to evolve the wave function into a pure state with no entanglement~\cite{chen_local_2010, wen_zoo_2017, nussinov_symmetry_2009, eisert_colloquium_2010} and the entanglement is thus called \textit{local}. 
Since such states belong to the same class as product states with zero entanglement, their entropy follows an area-law scaling with no offset ($\gamma=0$ in Eq.~\eqref{eq:area_law}). 
In contrast, for topologically ordered states, no such transformation exists (i.e.\ no local unitary can map the wave function into a pure product state with zero entropy), thus the entanglement is \textit{long range}. The absence of this local deformation results in a constant shift $-\gamma$ in the entanglement entropy, translating the impossibility of obtaining a zero-entropy state smoothly.

The TEE serves as a direct indicator of the specific kind of topological order one is faced with. The anyonic properties define the total quantum dimension $\mathcal{D} = \sqrt{\sum_a d_a^2}$, where $d_a$ denotes the local dimension of particles within the superselection sector $a$~\cite{kitaev_topological_2006, levin_detecting_2006}. The corresponding TEE then reads $\gamma = \ln(\mathcal{D})$. The present study focuses on characterizing a QSL with \Ztwo topological order, which exhibits four superselection sectors of abelian anyons. Consequently, it is characterized by $\gamma = \ln(\sqrt{4}) = \ln(2)$~\cite{hamma_bipartite_2005, nussinov_symmetry_2009, hamma_ground_2005}.

Various approaches have been proposed to extract the TEE numerically~\cite{kitaev_topological_2006, levin_detecting_2006}, where the general concept is to consider multiple domains in the lattice, either touching or overlapping. By subtracting the entropy values of the combined system from the entropy of the subregions, any dependence on boundary lengths, or even corner contributions, cancel out exactly, allowing for a direct determination of the TEE. 
We adopt the Kitaev-Preskill prescription~\cite{kitaev_topological_2006}, where three regions $A$, $B$ and $C$ converge at a triple intersection point. These regions collectively form a disk. The TEE is then expressed as the following linear combination
\begin{equation}
    -\gamma = S_A + S_B + S_C - S_{AB} - S_{BC} - S_{CA} + S_{ABC} \,\text{,}
\label{eq:TEE}
\end{equation}
where $S_{XY\dots}$ denotes the entanglement entropy of the composite region $X \cup Y \cup\dots$. This approach offers several advantages, including the absence of linear extrapolation to eliminate area-law terms. Additionally, it entails solely contractible entanglement boundaries (provided that no partition intersects a physical boundary), eliminating the reliance on the state decomposition (see Ref.~\cite{zhang_quasiparticle_2012}, where the TEE is contingent upon the minimal-entropy state decomposition of the wave function).

\section{\label{sec:model} Rydberg quantum simulator}
\begin{figure}[ht]
    \includegraphics[width=.6\linewidth]{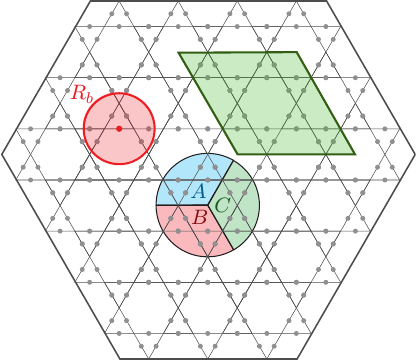}
    \caption{\label{fig:lattice} Lattice used in the experiment. The red circle represents the blockade radius $R_b$ set during the process. The green parallelogram describes the size of the lattice used for exact simulations in Sec.~\ref{sec:benchmark}. Subsets A, B, and C indicate the tripartition considered to estimate the topological entanglement entropy in Sec.~\ref{sec:entropy}.}
\end{figure}

We consider a physical system made of a two-dimensional array of Rydberg atoms placed at the edges of a Kagome lattice, as introduced in Ref.~\cite{semeghini_probing_2021} and depicted in Fig.~\ref{fig:lattice}. Each atom can either be in its electronic ground state $\ket{g}$ or excited to $\ket{r}$ via an optical transition with the Rabi frequency $\Omega_0$. Rydberg atoms interact through a repulsive van der Waals potential $V(r) = V_0 / r^6$ once excited~\cite{gelbart_van_2006, urban_observation_2009}. Due to this interaction, atoms closer than a characteristic distance $R_b$ cannot be excited at the same time. This phenomenon is known as \textit{Rydberg blockade} and the corresponding radius is defined as $R_b = (V_0/\Omega_0)^{1/6}$. The Hamiltonian for such a system is given by~\cite{saffman_quantum_2010, browaeys_manybody_2020}
\begin{equation}
    \hat{\mathcal{H}}(t) = -\frac{\Omega(t)}{2} \sum_i \hat{\sigma}_i^x - \Delta(t) \sum_i \hat{n}_i + \sum_{i<j} V(r_{ij}) \hat{n}_i \hat{n}_j \,\text{,}
 \label{eq:H}
\end{equation}
where $\hat{\sigma}_i^x = \dyad{r_i}{g_i} + \dyad{g_i}{r_i}$ and $\hat{n}_i = \dyad{r_i}$. Here, the Rabi frequency $\Omega(t)$ governs the strength of the coherent driving of the Rydberg transition whereas the time-dependent detuning $\Delta(t)$ serves as a chemical potential controlling the amount of excitations in the system.

In the following, we set the lattice spacing and Rabi frequency to respectively $a=\SI{3.9}{\um}$ and $\Omega_0=2\pi\times\SI{1.4}{\MHz}$, such that $R_b=2.4a$, corresponding to a blockade up to the third nearest neighbor, as illustrated in Fig.~\ref{fig:lattice}. Applied to the Kagome lattice with such settings, this constraint implies that at most one atom per triangle and per vertex is likely to be excited. Therefore, the physics of this system may be faithfully described as a dimer model. We identify~\cite{verresen_prediction_2021} an atom in its ground state with the absence of a dimer on that edge $\ket{g}=\ket{\monomer}$ and an atom in its excited state with the presence of a dimer $\ket{r}=\ket{\dimer}$, yielding, for example, $\ket{rgg} = \ket{\rgg}$ on a triangle.

\subsection{\label{sec:state_prep} Topological state preparation on a Rydberg device}

Recent studies~\cite{jahromi_topological_2020, verresen_prediction_2021, giudici_dynamical_2022, tarabunga_gaugetheoretic_2022, verresen_unifying_2022, samajdar_emergent_2023} have demonstrated that truncated versions of the Rydberg Hamiltonian~\eqref{eq:H} exhibit a QSL ground state within a certain range of the detuning parameter $\Delta$. Specifically, with exactly one dimer per vertex, the state manifests as a coherent superposition of compact dimerizations, akin to the RVB state~\cite{anderson_resonating_1973, rokhsar_superconductivity_1988, moessner_shortranged_2001}. 
However, since these simplified models only consider interactions up to the fourth neighbor, they entirely neglect the inherently long-range nature of the Rydberg interaction potential $V(r)$.
In this context, numerical studies~\cite[Supp. material ][]{semeghini_probing_2021, sahay_quantum_2023} have revealed that the presence of a long-range potential precludes the existence of a QSL ground state, and instead a direct transition to a valence bond \textit{solid} occurs. 
Notably, dynamical state preparation appears less sensitive to the presence of such tails of the potential, consistently yielding a QSL-like phase. 

In this context, a proper dynamical protocol compatible with a Rydberg device was proposed in Ref.~\cite{semeghini_probing_2021} for preparing and probing a topological spin liquid. This protocol involves slowly ramping up the frequencies $\Omega(t)$ and $\Delta(t)$ over time, allowing the system to start from a Hamiltonian whose ground state is easy to prepare and transition to the desired Hamiltonian of interest. 
First, the system is initialized in the state $\ket{\psi(t=0)} = \ket{g}^{\otimes N}$, which is the sole ground state of the Hamiltonian for $\Omega(t)=0$ and $\Delta(t)<0$. Subsequently, the state undergoes evolution under the time-dependent Hamiltonian in Eq.~\eqref{eq:H}. Both frequencies are then gradually increased (see App.~\ref{sec:App_protocol} for further details on the protocol used in this work) until reaching the value $\Delta>0$ for which $\ev{\hat{n}}\approx1/4$. 
It is crucial for this evolution to be faster than adiabatic to avoid remaining in the instantaneous ground state, where no QSL exists because of the long-range tails~\cite{semeghini_probing_2021, sahay_quantum_2023}. However, it must proceed at a pace slow enough as to avoid spurious excitations of higher-energy states~\cite{born_beweis_1928, kato_adiabatic_1950}. This leads the system to a \textit{metastable} state---a long-lived state despite not being the lowest eigenstate~\cite{gustafson_mathematical_2020, macieszczak_theory_2016, venuti_adiabaticity_2016}. As such, the speed of evolution must strike a delicate balance between adiabaticity and long-range stabilization in order to target the QSL phase.

\subsection{\label{sec:PQ} String Operators}

\begin{figure}[ht]
    \includegraphics[width=\linewidth]{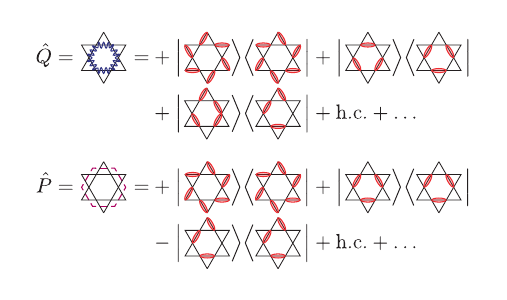}
    \caption{\label{fig:PD_def} Illustration of the action of both topological string operators $\hat{P}$ and $\hat{Q}$ defined on one hexagon. The red ellipses\protect\dimer represent the dimers on the lattice. 
    The operator $\hat{Q}$ maps a correct dimerization to another correct one by shuffling the dimers on the triangles it crosses with the wiggly line.
    The sign structure of the diagonal operator $\hat{P}$ is dictated by the parity of the number of defects encircled by the dashed line. 
    The representation is inspired by Ref.~\cite{verresen_prediction_2021}. }
\label{fig:PQ_def}
\end{figure}

Since the entanglement entropy of topologically ordered states cannot be measured on quantum hardware for relevant system sizes, one needs in practice to rely on different tools to identify the QSL phase. As anyonic properties are linked to the kind of topological order of the state, following Ref.~\cite{verresen_prediction_2021} we define in Fig.~\ref{fig:PQ_def} topological string operators mapped to the dimer representation~\cite{misguich_quantum_2002}. The non-diagonal operator $\hat{Q}$ acts on strings defined in the lattice by a wiggly line \QDrawing. When defined on a closed contour, it maps every valid dimerized configuration to another valid dimerization, thus measuring the coherence between them. 
The diagonal operator $\hat{P}$ acts on edges crossed by the dashed line \PDrawing and its parity on one single dimerization indicates whether defects are present or the dimerization is perfect. Thus, it quantifies the quality of each dimerization present in the state.
For the vacuum ground state of the \Ztwo topological order, i.e. the coherent superposition of perfect dimerizations, both operators evaluate to $1$. 

Moreover, both operators are associated with the creation of anyons when applied on an open string. Consequently, applying an open $\hat{Q}$ or $\hat{P}$ to the vacuum ground state will generate a pair of either $e$ or $m$-anyons, respectively, at the endpoints of the contour. As anyons are excitations of the system, this will result in projecting the ground state to an (orthogonal) excited state of the Hamiltonian. This feature may be exploited to probe the QSL. To this aim, we introduce the Fredenhagen-Marcu order parameters~\cite{fredenhagen_confinement_1986, bricmont_order_1983}, which correspond to the normalized application of topological operators on open contours:
\begin{equation}
    \QFMDrawing \quad \text{,} \quad \PFMDrawing \,\text{.}
\label{eq:FM}
\end{equation}
These operators must vanish for a QSL and thus constitute good order parameters. However, notice that the anyonic picture they rely upon is strictly correct only in the QSL phase~\cite{fradkin_phase_1979}. Furthermore, they become ill-defined outside of this region since the values of the $\hat{P}$ and $\hat{Q}$ loops in the normalization might vanish therein. They are thus relevant only within the QSL phase. 

\section{\label{sec:methods} Numerical Methods}

Considering a system of $N$ spins $1/2$ whose Hilbert space is spanned by $\{\ket{\uparrow}, \ket{\downarrow}\}^{\otimes N}$, we map the dimer representation to this space using the correspondence $\ket{g} = \ket{\uparrow}$ and $\ket{r} = \ket{\downarrow}$. In order to efficiently simulate the state preparation, we work in a restricted Hilbert space with at most one excitation per triangle $\hilb^\mathrm{restr} = \mathrm{Span}\left \{ \ket{\ggg}, \ket{\ggr}, \ket{\grg}, \ket{\rgg} \right \}^{\otimes N/3}$. 

\subsection{Variational model and its time evolution}
We approximate the wave function with the following variational Ansatz: 
\begin{equation}
    \psi_{\boldsymbol{\theta}}^{\text{JMF}}(\boldsymbol{\sigma}) = \exp \left( \sum_{i<j} \sigma_i V_{d_{ij}} \sigma_j \right) \prod_i \varphi_i(\sigma_i) \,\text{,}
    \label{eq:JMF}
\end{equation}
where $d_{ij}$ denotes the Euclidean distance between any two sites $i$ and $j$, and $\sigma_i \in \{\uparrow,\downarrow\}$ the spin of the atom at site $i$. Here, the mean field values $\varphi_i(\uparrow)$, $\varphi_i(\downarrow)$ as well as the Jastrow~\cite{jastrow_manybody_1955} potential $V_{d}$ play the role of variational parameters. We elaborate on the practical and physical grounds of this Ansatz in App.~\ref{sec:App_ansatz}. In particular, we show that this form is capable of exactly representing the RVB state. 

This state is evolved by varying its parameters $\boldsymbol{\theta}(t)$ in time using the time-dependent Variational Monte Carlo (t-VMC) prescription~\cite{carleo_localization_2012, carleo_unitary_2017}: 
\begin{equation}
    \dot{\theta}_k(t) = -i \sum_{k^\prime} (S^{-1})_{kk^\prime}  C_{k^\prime} \,\text{,}
\label{eq:TDVP}
\end{equation}
where the quantum geometric tensor $S$ and the vector of forces $C$ defined in Sec.~\ref{sec:App_MCMC} are evaluated using Monte Carlo integration. 

The Rényi-$2$ entanglement entropy of a given wave function $\psi$ is estimated by sampling two replicas of the system~\cite{zhao_measuring_2022, torlai_neuralnetwork_2018, hastings_measuring_2010}. For each replica, we partition the sites into two regions, $X=\{\ket{\uparrow},\ket{\downarrow}\}^{\otimes N_X}$ and its complement $Y=\{\ket{\uparrow},\ket{\downarrow}\}^{\otimes (N-N_X)}$, and evaluate
\begin{equation}
    S^{(2)}(\rho_X) = -\ln\left[ \mathbbm{E}_{\substack{\bm{\sigma} \sim |\psi|^2 \\ \bm{\sigma}^\prime \sim |\psi|^2 \\ }} \left [ \frac{\psi(\bm{\sigma}_X^\prime,\bm{\sigma}_Y)\psi(\bm{\sigma}_X,\bm{\sigma}_Y^\prime)}{\psi(\bm{\sigma}_X,\bm{\sigma}_Y)\psi(\bm{\sigma}_X^\prime,\bm{\sigma}_Y^\prime)} \right ]\right]\,\text{,}
\label{eq:renyi2}
\end{equation}
where $\bm{\sigma}_X, \bm{\sigma}_X^\prime \in X$ and $\bm{\sigma}_Y, \bm{\sigma}_Y^\prime\in Y$ are configurations in either region of the lattice and $\boldsymbol{\sigma}^{(\prime)} =(\bm{\sigma}_X^{(\prime)}, \bm{\sigma}_Y^{(\prime)}) \in \hilb$ form the basis states of the full Hilbert space. To estimate the TEE~\eqref{eq:TEE}, a tripartition as depicted in Fig.~\ref{fig:lattice} is chosen with partitions sharing boundaries and intersecting at a triple point, in adequacy with the constraints of the Kitaev-Preskill prescription. 

\subsection{\label{sec:benchmark} Benchmark of t-VMC}

The exact time-evolved state can be computed for a small lattice of $N=24$ sites with the geometry depicted in Fig.~\ref{fig:lattice}. Periodic boundary conditions are used to reduce finite-size effects. This exact solution allows us to benchmark the t-VMC method and assess whether it is accurate enough to describe our system. 
As shown in detail in App.~\ref{sec:App_expressivity}, the variational simulation closely matches the exact one from the beginning of time evolution. Up to $t=\SI{2.0}{\us}$, both simulations are practically indiscernible. At later times, as the system enters the dimerized regime, the qualitative behavior remains similar, even though a slight departure from the reference is observed. This shift is sufficiently small in all observables to conclude in a close match between both simulations. 

Through a fidelity analysis between the variational state and the exact solution at all times, we show in App.~\ref{sec:App_expressivity} that the t-VMC algorithm does not contribute significantly to the overall error between exact and variational results. Since this further rules out other numerical factors, such as the resolution of the equation of motion~\eqref{eq:TDVP} and Monte Carlo sampling, we conclude that the Ansatz is the sole significant source of the disparity. This allows us to conclude that our variational wave function, while able to represent the exact RVB state, also faithfully captures the state of the system at all times. As shown in App.~\ref{sec:App_expressivity}, this is in contrast to Ansätze previously proposed to investigate similar physics~\cite{giudici_dynamical_2022}, which prove unable to faithfully represent the state of the system at any time during the considered preparation protocol, and in particular the state at small times $t \leq \SI{2.0}{\us}$, where an infidelity higher than $10\%$ is found.

\section{\label{sec:results} Results}

In this section, we use our method to simulate the dynamical state preparation protocol of Sec.~\ref{sec:state_prep} from first principles. The variational results denote simulations conducted on the Rydberg atom lattice in Fig.~\ref{fig:lattice} with $N=219$ atoms, while exact simulations are obtained on a smaller lattice of $N=24$ Rydberg atoms, with periodic boundary conditions to minimize finite size effects. 

\subsection{\label{sec:dynamics} Time dependence of the topological operators}
\begin{figure*}[ht]
    \includegraphics[width=\linewidth]{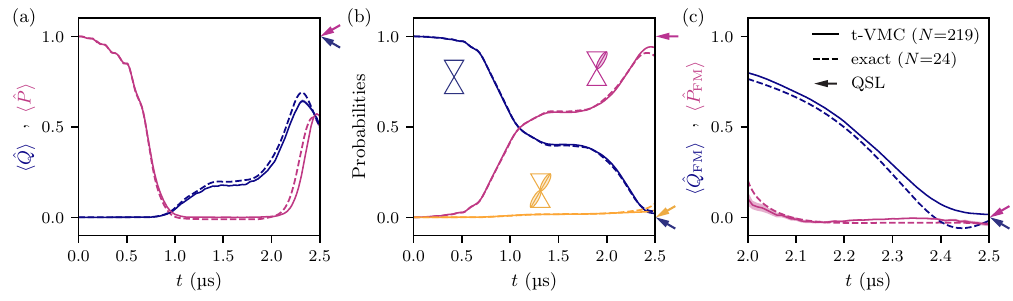}
    \caption{\label{fig:EE_vs_tVMC} Simulation of the dynamical topological state preparation. Variational results (t-VMC) using the JMF Ansatz on a Ruby lattice of $N=219$ atoms (as introduced in Fig.~\ref{fig:lattice}); statistical confidence intervals are narrower than the line width. Dashed lines indicate the exact simulation for a small lattice of $N=24$ atoms (as sketched in green in Fig.~\ref{fig:lattice}). Side arrows denote the expected values for an ideal QSL. Evolution of the (a) $\hat{P}$ (purple) and $\hat{Q}$ (dark blue) operators, as defined in Fig.~\ref{fig:PQ_def}, (b) probabilities to find a monomer (dark blue), dimer (purple) or double dimer (yellow) in the evolved state, and (c) the FM order parameters, as defined in Eq.~\eqref{eq:FM}. The FM order parameters are only computed at large times since their definition is only meaningful near the QSL phase, where the closed loops take finite values. All observables are averaged over all instances (vertices, hexagons, half-hexagons) present in the bulk of the lattice.}
\end{figure*}

During time evolution, we expect to go through a QSL-like phase, in which the state is characterized by a density of exactly one dimer per vertex. As discussed in Sec.~\ref{sec:PQ}, both topological operators presented in Fig.~\ref{fig:PQ_def} must have a finite expectation value whereas the FM order parameters in Eq.~\eqref{eq:FM} must vanish in the QSL state.

As all these observables are intensive quantities, it is possible to compare the results on various system sizes. In particular, we present in Fig.~\ref{fig:EE_vs_tVMC} both the variational results on the physical lattice ($N=219$) as well as exact evolution results on the smaller model ($N=24$), depicted in Fig.~\ref{fig:lattice}. 
We observe the onset of a QSL phase appearing after the trivial phase at $t=\SI{2.0}{\us}$, where all the observables change behavior and converge close to their expected values for a QSL. 
However, at no point do the topological operators coincide with their QSL predictions. Indeed, since a small number of monomers is always present in the bulk of the lattice as well as a small amount of double dimers, the value obtained for $\P$ is diminished as compared to the QSL case. In a complementary way, while $\Q$ measures the superposition and coherence of connected dimerizations, the observation $\Q<1$ implies that the state prepared is not an equal weight superposition like a QSL. 

On the other hand, while the FM order parameters start converging at the same time as other observables, they actually both reach the $0$ value expected for the QSL. Indeed, $\PFM$ is zero-valued throughout the whole dimerized regime $t>\SI{2.0}{\us}$, not providing any insightful information on the location of a QSL phase. In contrast, the off-diagonal order parameter reaches $\QFM \approx 0$ at $t>\SI{2.4}{\us}$, suggesting the QSL phase for $\Delta/\Omega_0\gtrsim 5.5$ equivalently. 
While this convergence to the expected RVB values suggests a QSL state, the significant discrepancy between the topological operator predictions and the QSL values highlights the inadequacy of the FM order parameters to detect a departure from an ideal QSL.

\subsubsection{\label{sec:noise} Effect of noise}
\begin{figure}[ht]
    \includegraphics[width=\linewidth]{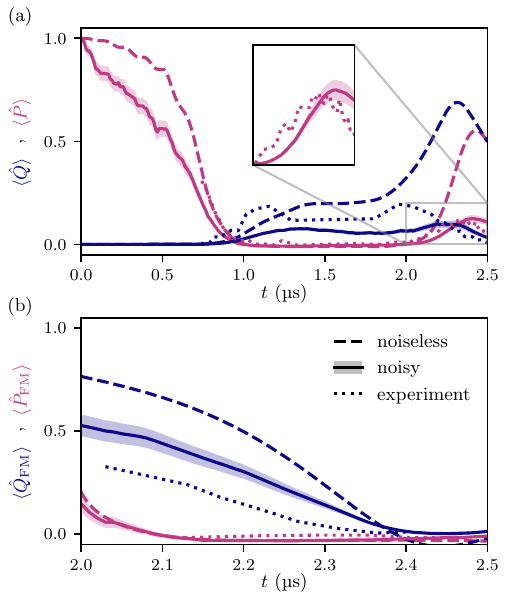}
    \caption{\label{fig:noise} Effect of noise on the dynamics of (a) the topological string operators and (b) FM order parameters for a small lattice of $N=24$ Rydberg atoms. The different colors indicate the measured operators. Different line styles denote different settings: noiseless exact evolution (dashed), noisy exact evolution (solid) and experimental measurements (dotted) from Ref.~\cite{semeghini_probing_2021} on a lattice of $N=219$ atoms. The results obtained for $\P$ are highlighted in the inset. The statistical confidence intervals are represented by lighter bands. The details of the choice of the noise parameters are discussed in App.~\ref{sec:App_noise}. }
\end{figure}

In this work, we are considering the dynamical preparation of a QSL state using a Rydberg Hamiltonian from first principles. However, a state prepared on a real-world quantum device might differ from numerical simulations non-trivially due to the spurious effects of noise throughout state preparation. 
Thus, we aim to qualitatively assess the effect of noise in a realistic quantum simulator and bridge the gap between numerical results in absence of nonidealities and experimental results.
For this purpose, we consider multiple error channels with their own error rates. Furthermore, we consider the possibility of spatial inhomogeneities of the fields $\Omega(t)$ and $\Delta(t)$. 
To take these effects into account, we use a scheme based on the stochastic unraveling of open quantum systems~\cite{plenio_quantumjump_1998}. The details on the algorithm and the effect of each individual source are discussed in detail in App.~\ref{sec:App_noise}.

The simulations in Fig.~\ref{fig:noise} show that in both noiseless and noisy simulations, the topological operators present a peak after $t=\SI{2}{\us}$, hinting at a transition to a topologically ordered phase. 
However, as thoroughly discussed in App.~\ref{sec:App_noise}, the addition of noise to the model has the effect of decreasing all expectation values throughout the evolution. For $\P$ and $\Q$ in particular, the diminishing of the peaks demonstrates that noise introduces more defects in the dimerizations and that connected dimerizations present less coherence. At short times, in the non-dimerized regime ($\ev{\hat{n}} \approx 0$), the behavior of $\P$ is significantly impacted by noise. 
Interestingly, the impact of noise on the FM order parameters is less pronounced than on the topological operators. In fact, while the measurements of topological operators in both open and closed contours are affected by noise, the value of $\Q$ is used as a normalization for $\QFM$. Thus, the reduction of amplitude is compensated for through the ratio of the two expectation values on the open and closed contour, respectively.

\subsubsection{Connection to experimental results\label{sec:conn_exp}}

To validate this analysis, we use the error rates presented in Ref.~\cite{semeghini_probing_2021} in order to compare our simulations with the experimental results. As shown in Fig.~\ref{fig:noise}, the measurements from the experiment are generally greatly diminished compared to our noiseless simulations. 
We observe a close agreement between the noisy simulations and the experiment for the diagonal operators $\P$ and $\PFM$, while the values for operators $\Q$ and $\QFM$ quantitatively differ. This discrepancy originates from the fact that these operators are not diagonal in the measurement basis. Hence, in the experiments, an additional protocol is conducted to rotate the state to the natural basis of these operators. This second quench uses modified frequencies $\tilde\Delta=0$ and $\tilde\Omega\neq\Omega_0$ to change the blockade radius of the Hamiltonian. In such a setting, the previously considered noise channels have modified rates while additional decoherence channels might appear. Furthermore, as the blockade radius decreases, states containing violations of the original constraint, such as vertices with double dimers, might be altered. Since these effects are absent from our rather predictive simulations, one can quantify the significant additional error induced when evaluating experimentally off-diagonal observables.


\subsection{\label{sec:logic} Logical operations}

\begin{figure*}
    \includegraphics[width=\linewidth]{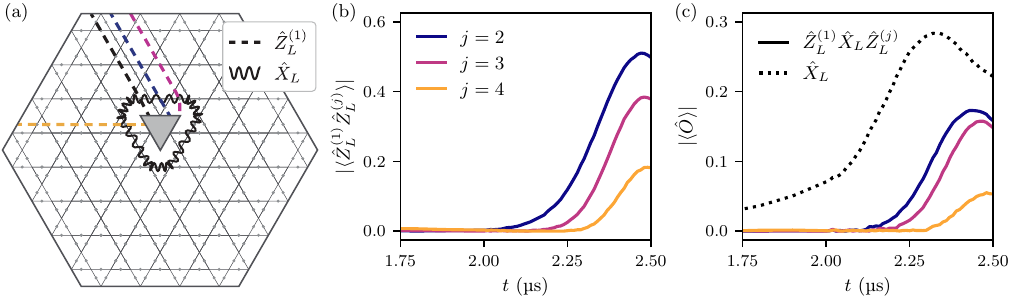}
    \caption{\label{fig:logic}
    (a) Schematic illustration of the pierced lattice with $N=285$ Rydberg atoms and definition of the logical operators. The filled gray triangle represents a hole in the lattice where no atom is present, and therefore acts as an additional boundary. The logical operator $\hat{X}_L$ (black wiggly line) acts on a string looping around the central hole and $\hat{Z}_L$ (dashed lines) acts on sites connecting the hole to the external boundary. 
    The action of both operators on the logical states is described in Eq.~\eqref{eq:logical_operators}.
    We define multiple operators $\hat{Z}_L^{(j)}$ corresponding to various choices of topologically equivalent paths.
    (b) Measurements of the Pauli identity $\hat{\mathbbm{1}} = \hat{Z}_L \times \hat{Z}_L$ for various pairs of logical operators. As the sign of the expectation value depends on the number of enclosed vertices, the absolute value is plotted here. 
    (c) Verification of the identity $\hat{X}_L = - \hat{Z}_L \times \hat{X}_L \times \hat{Z}_L$ for our logical operators. As for (b), we take the absolute value due to the dependence of the sign on the number of enclosed vertices. 
    }
\end{figure*}

We verified in Sec.~\ref{sec:dynamics} that the prepared state possesses characteristics consistent with topological order found in the surface code. A further salient characteristic of such a code is the condensation of anyons at the physical boundaries of the lattice. For this reason, the topology of the simulated system has a crucial impact on this phenomenon. We here investigate this by adding a hole in the bulk of the planar lattice, thereby creating a second physical boundary~\cite{semeghini_probing_2021, verresen_prediction_2021}. This induces the condensation of $m$ anyons on both the external boundary and the hole, separating the Hilbert space of dimer coverings into two topological sectors, which can be labeled as \textit{logical} quantum bits $\ket{0}_L$ and $\ket{1}_L$, as introduced by Ref.~\cite{semeghini_probing_2021}. While this approach poses severe challenges for DMRG-based methods due to the shape of the lattice, the flexibility of our method allows for a complete study of this system. By simulating a large system of $N=285$ atoms we can predict the topological entropy in the presence of a bulk (see Sec.~\ref{sec:entropy} and App.~\ref{sec:App_TEE}).

\subsubsection{Definition}
In this framework, we define \textit{logical operators} $\hat{X}_L$ and $\hat{Z}_L$ acting on the perfect logical states $\ket{0}_L$ and $\ket{1}_L$,
\begin{equation}
\begin{aligned}
    \hat{X}_L\ket{0}_L &= \ket{1}_L, & \hat{Z}_L\ket{0}_L &= \ket{0}_L, \\
    \hat{X}_L\ket{1}_L &= \ket{0}_L, & \hat{Z}_L\ket{1}_L &= -\ket{1}_L,
\label{eq:logical_operators}
\end{aligned}
\end{equation}
on paths illustrated in Fig.~\ref{fig:logic}(a). The off-diagonal operator $\hat{X}_L$ maps one logical state to the other and is applied to a path that circumvents the hole. The diagonal operator $\hat{Z}_L$ allows for the identification of the spin projection of the logical qubit and is applied to a path that connects the two separated boundaries. 
The definition of both logical operators roots in the anyonic properties of the vacuum ground state of the toric code introduced in Sec.~\ref{sec:PQ}. In the anyonic picture, $\hat{Z}_L$ acts by connecting $m$-anyons on both boundaries together. Also, $\hat{X}_L$ acts by creating an $e$-anyon, winding it around the $m$-anyons condensed on the hole, and finally destroying it. Therefore, both operators can be defined in multiple ways, as long as the topology of the path they are applied on remains unchanged.

Locally, these logical operators act the same way as the previously defined topological operators $\hat{Q}$ and $\hat{P}$. As a consequence, the sign of the expectation value of any closed diagonal operator will solely be dictated by the parity of the number of vertices enclosed by its path, namely $(-1)^{\# \text{vertices}}$~\cite{verresen_prediction_2021}. 
On the other hand, we showed in previous analyses that both $\P$ and $\Q$ reached reduced values compared to a perfect QSL in the simulation of the Rydberg experiment. As the magnitude of topological operators is affected by the presence of defects, we expect the same behavior to occur for the logical operators. In that case, the previously stated properties of the logical operators $\hat{X}_L$ and $\hat{Z}_L$ are only approximate and the exact impact on the results is to be determined.

Equipped with these operators, it is possible to identify the state prepared on the logical qubit. We find $\langle\hat{X}_L\rangle > 0$ and $\langle\hat{Z}_L\rangle \sim 0$ in the approximate dimer phase, identifying our state with an approximate logical state $\ket{\psi} \sim \ket{+}_L = (\ket{0}_L + \ket{1}_L)/\sqrt{2}$ with additional defects. Despite the topology being different from the lattice studied in Sec.~\ref{sec:dynamics}, the prepared state is also an approximate superposition of dimerizations from both topological sectors, thus approaching the RVB state in the same fashion as for the planar lattice without a hole defined in Fig.~\ref{fig:lattice}. Moreover, the condensation of $m$-anyons at both boundaries is confirmed by the expectation values of both logical operators.

\subsubsection{Pseudo-identity}

We further use the Pauli identity $\hat{\mathbbm{1}} = \hat{Z}_L \times \hat{Z}_L$ that holds for perfect topological qubits and detect deviations from this identity to characterize the imperfect QSL.
In a state with perfect topological invariance of the logical operators, the identity holds more generally (up to a sign) when we consider the product of two different logical operators $\hat{Z}_L^{(1)} \hat{Z}_L^{(j)}$.
However, the parity is non-trivial and depends entirely on the number of vertices enclosed between the two strings through $(-1)^{\# \text{vertices}}$. We will therefore evaluate the identity at increasing separation distance between the two strings on which the $\hat{Z}_L^{(j)}$ operators are defined. 

We numerically verified the expected parity sign structure. We show the absolute values of the desired identity for multiple pairs of paths that make up loops that contain an increasing number of sites in Fig.~\ref{fig:logic}(b). 
The qualitative behavior is similar for all loops and confirms a finite expectation value at late times, when a QSL-like phase was diagnosed in the dynamic state preparation. However, a significant difference appears in the magnitude of the different observables. We observe that, as the two operators $\hat{Z}_L^{(1)}$ and $\hat{Z}_L^{(j)}$ are further away from each other,  $\abs{\langle\hat{Z}_L^{(1)} \hat{Z}_L^{(j)}\rangle}$ decreases. This observation confirms that, even though the prepared state presents local features of a QSL, it contains non-trivial differences from an RVB state that clearly appear in larger-scale observables.

\subsubsection{Pauli algebra}

As we only have access to the dynamically prepared state $\ket{\psi}\sim\ket{+}_L$, it is not possible to fully characterize the logical algebra we defined.
However, as a first step to verify the structure of Lie algebra, we evaluate the identity $\{ \hat{X}_L,\hat{Z}_L\} = 0 \iff \hat{X}_L = -\hat{Z}_L \times\hat{X}_L \times\hat{Z}_L$. In experiments, the verification of this identity poses serious challenges, since the observables to be observed act on a large number of sites and consist of both diagonal and non-diagonal operators. As shown in Sec.~\ref{sec:conn_exp}, the interplay between experimental noise and the basis rotation involved in the measurement of the latter yields sizable spurious effects.

As in the study of the identity in the previous subsection, the use of a closed loop composed of two separate $\hat{Z}_L$ is required for the operator to be non-trivial. Here also, a parity factor appears that depends on the number of vertices enclosed between the $\hat{Z}_L$ operators. We verified that the sign of $\langle \hat{Z}_L^{(1)} \hat{X}_L \hat{Z}_L^{(j)} \rangle$ corresponds to $-(-1)^{\# \text{vertices}}$, and thus is exactly opposite to the sign of the pseudo identity $\langle \hat{Z}_L^{(1)} \hat{Z}_L^{(j)} \rangle$. 
Given that the sign of $\ev{\smash{\hat{X}_L}}$ is always positive, we show the absolute value of the rotated $\hat{X}_L$ in Fig.~\ref{fig:logic}(c). 
Before the onset of the topologically ordered phase around $t = \SI{2.2}{\us}$, we observe that $\lvert\ev{\smash{\hat{X}_L}}\rvert$ and its rotated counterpart $\lvert\langle\hat{Z}_L^{(1)} \hat{X}_L \hat{Z}_L^{(j)}\rangle\rvert$ qualitatively differ. 
In contrast, at later times when the system exhibits topological order, the qualitative behavior of all curves agrees for all realizations of the logical operator $\hat{Z}_L$ defined on any of the tested topologically equivalent paths. We find that the further apart the operators $\hat{Z}_L^{1}$ and $\hat{Z}_L^{j}$ are, the poorer the agreement.
This allows us to confirm that the operators approximately respect the Pauli algebra at small distances and thus can be considered appropriate logical operators.
Deviations in the expectation value of these observables are again due to defects.

The analysis conducted in this section is a striking illustration of the strengths of our method, namely ($i$) its flexibility in terms of geometry and topology, and ($ii$) its ability to directly measure string operators, whether diagonal or not. 
A thorough study of the anyonic properties of topologically ordered states by evaluating products of non-commuting string operators should entail a prohibitive overhead for experiments, which would require subsequent local-basis rotation.

\subsection{\label{sec:entropy} Entanglement entropy}
\begin{figure}[ht]
    \includegraphics[width=\linewidth]{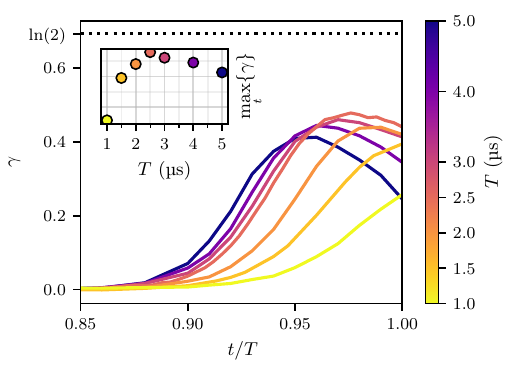}
    \caption{\label{fig:gamma} Dynamics of the topological entanglement entropy for various total preparation times $T$. The dotted line indicates the value obtained for the RVB form of the state~\eqref{eq:App_RVB}, corresponding to $\gamma = \ln(2)$. The inset displays the highest $\gamma$ reached for each simulation. The simulation with total time $T=\SI{2.5}{\us}$ corresponds to the experimental protocol from~\cite{semeghini_probing_2021}. TEEs and their errors were obtained from $1024$ bootstrap estimates~\cite{gao_estimating_2008}.}
\end{figure}

As discussed in Sec.~\ref{sec:dynamics}, while the FM order parameters are good indicators of QSL-like behavior, they are ill-defined outside of this regime. Furthermore, the values reached by the string operators $\hat{P}$ and $\hat{Q}$ were found to exhibit a substantial dependence on the volume enclosed by the loop~\cite{verresen_prediction_2021}, raising questions about the existence of a genuine long-range order extending across the entire system. This issue can be addressed through the value of the topological entanglement entropy (see Sec.~\ref{sec:entropy}), which can unambiguously confirm the presence of topological order and its nature.

We time evolve the system of $N=219$ Rydberg atoms according to the dynamical preparation protocol introduced in Ref.~\cite{semeghini_probing_2021} (see App.~\ref{sec:App_protocol} for details on the protocol), matching its geometry, topology and system size, and for various preparation times, in particular that of the experiment $T=\SI{2.5}{\us}$. Using Eq.~\eqref{eq:renyi2}, we estimate the entanglement entropy and further measure the TEE using the Kitaev-Preskill prescription~\cite{kitaev_topological_2006} as defined in Eq.~\eqref{eq:TEE} for the tripartititon depicted in Fig.~\ref{fig:lattice}. The results in Fig.~\ref{fig:gamma} show no TEE for most of the time evolution, up to $t = 0.9 T = \SI{2.25}{\us}$, confirming the absence of topological order in the trivial phase at low $\Delta$. At later times, the system transitions to a finite TEE for $t \gtrsim \SI{2.25}{\us}$, attaining a maximal value of $\gamma \approx 0.479(2)$ at $t \approx \SI{2.44}{\us}$, corresponding to $\Delta/\Omega_0 \approx 6$. This confirms the presence of topological order in the prepared state. However, we also observe that the system never reaches the characteristic value $\gamma=\ln(2)$ corresponding to \Ztwo topological order. Thus, the final state is not a pure QSL.

As the topology of the lattice can ultimately affect the final state, we study in App.~\ref{sec:App_TEE} its influence on the TEE. Indeed, as previous studies considered lattices with different geometries (cylinder or torus but no lattice with open boundaries), our usage of a lattice with open boundary conditions, as relevant in experiments, could have affected the topological order of the system. We thus verify that the topology of the lattice (boundary condition, genus, size) is not a significant source of the disparity between the TEE of the prepared state and that of an RVB. In particular, none of the alternative lattices contributed to an improvement in the value of $\gamma$.

As discussed in Sec.~\ref{sec:state_prep}, it has been shown~\cite{verresen_prediction_2021, semeghini_probing_2021, giudici_dynamical_2022} that systems analogous to the one simulated here, but with a finite range of the Rydberg potential, possess a QSL ground state for some values of the parameter $\Delta$. Therefore, for such systems, a perfectly adiabatic preparation $T\to\infty$ should ensure the maximal value of $\gamma=\ln(2)$. 
However, the case presented here differs from previous analyses by considering the complete physical Hamiltonian~\eqref{eq:H}, where the presence of \emph{long-range} interactions destabilizes the QSL~\cite{verresen_prediction_2021, giudici_dynamical_2022, sahay_quantum_2023}. Hence, as discussed in Sec.~\ref{sec:TEE}, increasing the total evolution time does not guarantee a higher TEE. Instead, there should be an optimal preparation time $T$ for which the final state is as close as possible from the targeted QSL.

Since the TEE provides a precise global indicator of topological order, in Fig.~\ref{fig:gamma} we assess its dependence on adiabaticity by comparing the obtained $\gamma$ as a function of the evolution time $T$ of the protocol. 
We find a finite TEE for $t \gtrsim 0.9 T$ that ensures the presence of topological order across all examined cases. We observe that for larger $T$ values, the peak of $\gamma$ occurs before the end of the quench $t=T$. Conversely, as $T$ decreases, the peak in $\gamma$ shifts rightwards. At small $T$, $\gamma$ does not reach an optimum. 

In Fig.~\ref{fig:gamma}, we also report the maximum value of $\gamma$ across different preparation times $T$ and identify that the preparation time that optimizes the TEE is $T=\SI{2.5}{\us}$ for the lattice geometry depicted in Fig.~\ref{fig:lattice}. This $T$ corresponds to the one used in Ref.~\cite{semeghini_probing_2021}, yielding $\gamma \approx 0.479(2)$. 
Even though a finite topological entanglement entropy ensures the presence of topological order for all examined cases, we observe that no state possesses the signature of \Ztwo topological order, since $\gamma<\ln(2)$ for all $T$. We thus conclude that the simulated protocol does not lead to the preparation of a pure RVB state.

\section{\label{sec:conclusion} Conclusion}

In this work, we propose time-dependent variational Monte Carlo as a method to efficiently simulate quantum many-body systems exhibiting topological order. Unlike previous state-of-the-art numerical studies, our approach can accurately describe the state throughout the whole dynamical protocol. Furthermore, it allows for the consideration of the physical Hamiltonian from first principles and is not limited by entanglement or short-ranged interactions, while also offering a total degree of control on the geometry and topology of the physical system. The versatility of the method---along with the linear scaling of the number of variational parameters with system size---allows for matching physically relevant lattice geometries at scale while giving direct access to all observables and to entanglement metrics. 

By simulating the protocol for the dynamical preparation of a quantum spin liquid on an array of interacting Rydberg atoms, we show that not only is our approach able to capture the onset of a quantum spin liquid phase, but also to faithfully represent the state of the system at all stages of the protocol. Moreover, we simulate the dynamics of lattices with up to $288$ atoms with multiple choices of boundary conditions and both genus $0$ and $1$, thereby demonstrating the flexibility of the method. 
Our numerical approach gives access to various observables to confirm the presence of topological order, such as the topological operators and the Fredenhagen-Marcu order parameters upon time evolution. It enables us to assess how the prepared state lies within the stabilizer space of a surface code by probing a representation of the Pauli algebra in terms of topological operators. Furthermore, we are able to efficiently measure the topological entanglement entropy throughout time evolution. 
As the former observables suggest properties akin to a QSL at the end of the preparation protocol, the latter allows to unequivocally confirm the presence of topological order. However, as the topological entanglement entropy differs from the signature value of a \Ztwo QSL, we show that the state never coincides with the expected RVB state. 

The use of numerical simulations for the optimization of the design of experimental protocols is a topic of renewed interest~\cite{brady_optimal_2021, pikulin_protocol_2021, giudici_dynamical_2022}. We foresee that the ability of our variational approach to faithfully match physical systems with long-range interactions as well as geometries relevant to experiments should make the present technique the method of choice in this context.
Another very promising prospect is quantum state tomography with variational Ansätze~\cite{torlai_neuralnetwork_2018, neugebauer_neuralnetwork_2020, schmale_efficient_2022}, where a variational state is optimized to best mimic the distribution of outcomes measured in real-world experiments. This approach then allows one to numerically estimate other quantities not directly accessible on the physical setup, such as entanglement entropy. The remarkable accuracy of our variational Ansatz in representing the wave function of physically relevant systems should make it a privileged candidate in this matter. 
Finally, all the presented methods straightforwardly extend to more accurate and expressive variational wave functions, such as neural quantum states, for which time dynamics has driven active research over the past years~\cite{schmitt_quantum_2020, sinibaldi_unbiasing_2023, nys_abinitio_2024}.

\begin{acknowledgments}
The simulations in this work were carried out using NetKet~\cite{vicentini_netket_2022, carleo_netket_2019}, which is based on Jax~\cite{bradbury_jax_2018} and MPI4Jax~\cite{hafner_mpi4jax_2021}.
This research was supported by a MARVEL INSPIRE Potentials Master's Fellowship, by the NCCR MARVEL, a National Centre of Competence in Research, funded by the Swiss National Science Foundation (grant number 205602), and by SEFRI through Grant No. MB22.00051 (NEQS - Neural Quantum Simulation).
\end{acknowledgments}

\appendix
\numberwithin{table}{section}
\numberwithin{figure}{section}
\section{\label{sec:App_QMC} Variational Monte Carlo}

The Rydberg blockade is strongest for first-neighbor interactions. Indeed, for distances smaller than $R_b$, the potential takes values $V(\firstnn)/\Omega_0\approx191$, $V(\secondnn)/\Omega_0\approx7$ and $V(\thirdnn)/\Omega_0\approx3$. Since the wave function of states close to the ground state is expected to have vanishing amplitudes for such configurations due to their high energy contributions, we effectively work in a constrained space, as introduced in~\cite{verresen_prediction_2021}. The usage of a total first-nearest-neighbor blockade has the advantage of reducing the number of states of the considered Hilbert space, as well as discarding high-frequency terms in the Hamiltonian, adding stability to the whole numerical scheme. 

\subsection{\label{sec:App_ansatz} Ansatz}
Upon considering the variational Ansatz described in Eq.~\eqref{eq:JMF}, the time dependence of the variational parameters $\boldsymbol{\theta}(t)$ completely captures the dynamics of the system throughout its evolution. Since the van der Waals interaction potential depends exclusively on the distance between sites, the Jastrow factor was chosen to be invariant in the same way. This choice of parametrization has the advantage of requiring a reduced number of parameters, scaling only linearly with the system size. We find that this Ansatz is a good compromise between expressivity and the induced computational burden of time evolution. Indeed, our ansatz does not impose any restrictions on the length of the correlations that can be captured by the model and is therefore compatible with the topological character of a QSL. More specifically, our Jastrow parametrization can faithfully represent an RVB state,
\begin{equation}
\begin{aligned}
    \ket{\text{RVB}} \propto& \prod_{v \in \mathcal{V}} \left( \prod_{i\in v} \hat{n}_i \prod\limits_{\substack{j\in v \\ j\neq i}} (1-\hat{n}_j) \right) \ket{+}^{\otimes N} \\
    \propto& \lim_{W\to -\infty} \exp\left[ W \sum_{i<j} \hat{n}_i \chi_{ij} \hat{n}_j - W\sum_i \hat{n}_i \right] \ket{+}^{\otimes N} \\
    \propto& \lim_{W\to -\infty} \exp \left[ \frac{W}{4}\sum_{i<j} \hat{\sigma}_i^z \chi_{ij} \hat{\sigma}_j^z \right] \\
    &\qquad\quad\times \exp\left[ - \frac{W}{2}\sum_i (z_i-1)\hat{\sigma}_i^z \right] \ket{+}^{\otimes N} \,\text{,}
\label{eq:App_RVB}
\end{aligned}
\end{equation}
where $\ket{+} = \left(\ket{\uparrow}+\ket{\downarrow}\right)/2$ is the equal-weight superposition of all basis states in the computational basis, $\mathcal{V}$ denotes the set of vertices of the lattice and $i \in v$ refers to all the atoms/sites $i$ connected to the same vertex $v$, $\chi_{ij} \in \{0,1\}$ indicates if the two sites $i$ and $j$ are connected to the same vertex, and $z_i=\sum_j \chi_{ij}$ is the number of sites sharing a vertex with $i$. While the limiting process can seem ill-defined whenever $\sigma^z_i \chi_{ij} \sigma^z_j = -1$, we highlight the fact that the normalization of the wave function, neglected here, lifts such concerns. 
The final expression has a two-body Jastrow form fully compatible with our Ansatz \eqref{eq:JMF}.

\subsection{\label{sec:App_MCMC} Monte Carlo estimates}
The Born distribution of the wave function $p(\boldsymbol{\sigma}) = |\psi_{\boldsymbol{\theta}}(\boldsymbol{\sigma})|^2/\braket{\psi_{\boldsymbol{\theta}}}$ can be sampled using Monte Carlo Markov Chain (MCMC). Along with this, expectation value of a local or sparse operator $\hat{O}$ is efficiently approximated using Monte Carlo integration $\ev*{\hat{O}} = \mathbbm{E}_{\vb*{\sigma}\sim p} \left [ O_\text{loc}(\boldsymbol{\sigma})\right ]$, which is an average of the local operator $O_\text{loc}(\vb*{\sigma}) = \sum_{\vb*{\sigma}^\prime} \mel{\vb*{\sigma}}{\hat{O}}{\vb*{\sigma}^\prime} \frac{\psi_{\boldsymbol{\theta}}(\vb*{\sigma}^\prime)}{\psi_{\boldsymbol{\theta}}(\vb*{\sigma})}$, where the sum runs over the few configurations $\vb*{\sigma}'$ for which the matrix elements $\mel{\vb*{\sigma}}{\hat{O}}{\vb*{\sigma}^\prime}$ are non-zero.

The time-dependent variational principle (TDVP)~\cite{carleo_localization_2012, yuan_theory_2019} equation of motion for the parameters $\boldsymbol{\theta}$ is obtained through the minimization of the Fubini distance between a state with new parameters $\ket{\smash{\psi_{\boldsymbol{\theta}(t)+\boldsymbol{\dot{\theta}}\delta t}}}$ and the evolved state $U(\delta t)\ket{\psi_{\boldsymbol{\theta}(t)}}$. The resulting equation~\eqref{eq:TDVP} depends on two quantities that can be sampled using Monte Carlo. Firstly, the quantum geometric tensor (QGT) describes the covariance of the derivatives of the wave function:
\begin{equation}
\begin{aligned}
    S_{ij}&= \frac{\bra{\d_i \psi}\ket{\d_j \psi}}{\braket{\psi}} - \frac{\bra{\d_i \psi}\ket{\psi}\bra{\psi}\ket{\d_j \psi}}{\braket{\psi}^2} \\
    &= \mathbbm{E}\left [ D_i^*(D_j - \mathbbm{E}[D_j]) \right ] \,\text{,}
\label{eq:S}
\end{aligned}
\end{equation}
where $D_k(\boldsymbol{\sigma}) = \partial_k \log \psi(\boldsymbol{\sigma})$ can be obtained by automatic differentiation~\cite{bradbury_jax_2018}. Secondly, the vector of forces represents the derivatives of the energy in the same parameter space:
\begin{equation}
\begin{aligned}
    C_i &= \frac{\bra{\d_i \psi} \hat{\mathcal{H}} \ket{\psi}}{\braket{\psi}} - \frac{\bra{\d_i \psi} \ket{\psi}\bra{\psi} \hat{\mathcal{H}}  \ket{\psi}}{\braket{\psi}^2}  \\
    &= \mathbbm{E} [ D_i^* (E_\text{loc} - \mathbbm{E}[E_\text{loc}] )] \,\text{.}
\label{eq:C}
\end{aligned}
\end{equation}
where we introduced the local energy $E_\text{loc}$.

\section{\label{sec:App_TDVP_MF} Mean-field evolution}
As recently shown~\cite{sinibaldi_unbiasing_2023}, the Monte Carlo estimators defined in Eq.~\eqref{eq:S} and ~\eqref{eq:C} are significantly biased whenever the amplitude of the variational wave function vanishes for basis states yielding finite contributions to the gradients. This is a generic feature for distributions with many zeros, as is the initial state in our case. To circumvent this issue, we evolve the initial state by analytically solving the TDVP equations, i.e. dispensing with any sampling, on a simplified Ansatz with all Jastrow parameters set exactly to zero: 
\begin{equation}
    \ket{\psi_\theta} = \bigotimes_{i=1}^N \left( \alpha_i \ket{\uparrow}_i + \beta_i \ket{\downarrow}_i \right) \,\text{,}
    \label{app:Ansatz}
\end{equation}
where we use the correspondence $\ket{g}=\ket{\uparrow}$ and $\ket{r}=\ket{\downarrow}$ and consider the parameters normalized $|\alpha|^2+|\beta|^2 = 1$.
The product-state nature of the Ansatz implies $S$ to be block-diagonal, only coupling the parameters $\beta_k$, $\alpha_k$ acting on the same site. Thus, we only need to solve a two-dimensional linear equation. Upon neglecting the indices $k$ and time $t$ for the forces to simplify notations, and defining $C_{\beta} = B$ and $C_{\alpha}=A$, one obtains the system:
\begin{equation}
\begin{aligned}
    \mqty(|\alpha|^2 & -\beta\alpha^* \\ -\beta^*\alpha & |\beta|^2) \mqty( \dot{\beta} \\ \dot{\alpha}) &= -i \mqty(B \\ A) \,\text{.}
\end{aligned}
\label{app:linear_system}
\end{equation}
To decrease the number of degrees of freedom, we fix the global phase by setting $\alpha^\text{I} = 0$, where superscripts R and I indicate real or imaginary part, respectively, and use the normalization constraint to express $\alpha$ and $\dot{\alpha}$ in terms of the other two parameters. The system \eqref{app:linear_system} can then be solved by diagonalizing $S$, replacing $\dot{\alpha}^\text{R}$ and decomposing into real and imaginary parts:
\begin{widetext}
\begin{equation}
\label{app:beta_dot_of_C}
\begin{aligned}
    \dot{\beta}^\text{R} = +((\alpha^\text{R})^2+(\beta^\text{I~})^2) B^\text{I~} + \beta^\text{I~}\beta^\text{R}B^\text{R} - \alpha^\text{R}\beta^\text{R}A^\text{I~} - \frac{1}{\alpha^\text{R}}\beta^\text{I~} A^\text{R} \,\text{,} \\
    \dot{\beta}^\text{I~} = - ((\alpha^\text{R})^2+(\beta^\text{R})^2)B^\text{R} -\beta^\text{I~}\beta^\text{R} B^\text{I~} - \alpha^\text{R}\beta^\text{I~}A^\text{I~} + \frac{1}{\alpha^\text{R}}\beta^\text{R} A^\text{R} \,\text{.} 
\end{aligned}
\end{equation}
\end{widetext}
Notice that in these equations, the division could be ill defined if $\alpha^\text{R}=0$. Fortunately, this will never be the case in our situation, since $(\alpha^\text{R})^2$ is the probability of having the site $k$ in $\ket{g}$, which is always close to $1$ for $\Delta < 0$ (in practice, this is true throughout the whole evolution).

We find that the particles are effectively subject to a local potential $v = -\Delta + \Omega\sum_{l\neq k}\bigl(\frac{R_b}{r_{kl}}\bigr)^6 |\beta_l|^2$. Then, the forces are given by 
\begin{equation}
\begin{aligned}
\label{app:C}
    B &= -\frac{\Omega}{2}\left(\alpha-2\beta\beta^\text{R}\alpha \right) + v \beta |\alpha|^2 \,\text{,} \\
    A &= -\frac{\Omega}{2}\left(\beta-2\alpha\beta^\text{R}\alpha \right) - v \alpha |\beta|^2 \,\text{,}
\end{aligned}
\end{equation}
where we evaluate the frequencies at the same time step. Putting everything together, we obtain
\begin{equation}
\begin{aligned}
    \dot{\beta}^\text{R} &= +v \beta^\text{I} - \frac{\Omega}{2} (-\frac{\beta^\text{I}\beta^\text{R}}{\alpha}) \,\text{,} \\
    \dot{\beta}^\text{I} &= -v \beta^\text{R} - \frac{\Omega}{2} (\frac{(\beta^\text{R})^2}{\alpha} -\alpha) \,\text{.} \\
\end{aligned}
\label{app:beta_dot}
\end{equation}

Finally, using normalization again and gathering $\beta = \beta^\text{R} + i\beta^\text{I}$, we obtain the final solution for the complex parameters:
\begin{equation}
\begin{aligned}
    \dot{\beta}_k(t) &= -i v_k(t) \beta_k - i\frac{\Omega(t)}{2} (\beta_k\frac{\beta_k^\text{R}}{\alpha_k} -\alpha_k) \,\text{,} \\
    \dot{\alpha}_k(t) &= -\frac{\Omega(t)}{2} \beta_k^I \,\text{.}
    \label{eq:TDVP_MF}
\end{aligned}
\end{equation}
This analytical evolution is conducted for a short starting time $t^* < T$, up to a point where the distribution has spread over more basis states and one can use the MCMC estimate of the TDVP equation. Typically, we set this value to $t^* = \frac{2}{25}T$, corresponding to $40\%$ of the time used for the sweeping of the Rabi frequency (see Sec.~\ref{sec:App_protocol}).

\section{\label{sec:App_protocol} Influence of the protocol on the results}

\begin{figure*}
    \includegraphics[width=\linewidth]{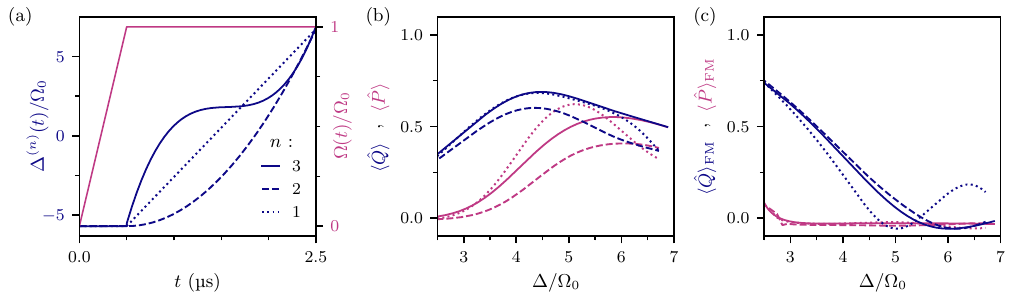}
    \caption{\label{fig:app_schedule} Comparison of different protocols through exact evolution of a small lattice of $N=24$ atoms. 
    (a) Time dependence of the parameters of the Hamiltonian for all considered protocols. All start with $\Omega(0)=0$ and $\Delta(0) = -2\pi\times\SI{8}{\MHz}$ and take place over a total time $T=\SI{2.5}{\us}$. The Rabi frequency $\Omega(t)$ is first linearly increased up to the target value $\Omega_0 = 2\pi\times\SI{1.4}{\MHz}$ during $t=T/5$, while the detuning is kept constant. The detuning is then ramped up following a polynomial of degree $n$ up to its final value $\Delta(T)=2\pi\times\SI{9.4}{\MHz}$. The protocol used in the main text uses $n=3$ and corresponds to that of Ref.~\cite{semeghini_probing_2021}. 
    (b) Expectation value of the topological string operators $\hat{Q}$ and $\hat{P}$ for the various forms of $\Delta(t)$ upon time evolution. 
    (c) Expectation value of $\QFM$ and $\PFM$ for the same simulations.}
\end{figure*}

As discussed in Sec.~\ref{sec:state_prep}, to approach the QSL state, a delicate balance of adiabaticity must be used in the protocol of preparation. Here, we will study the influence of the choice of protocol on the type of states reached during the state preparation. In this respect, we consider a family of similar protocols as depicted on Fig.~\ref{fig:app_schedule}, where the detuning increases as a polynomial function of the time after the Rabi frequency $\Omega(t)$ has attained its maximal value $\Omega_0$. While an unlimited amount of different protocols could be considered, through the modification of the initial and final values of both frequencies, the total sweeping time or the functional shape of each sweep, we are here only interested in ruling-out the protocol as the source for the lack of topological order by considering similar schedules. 

At small times, we observe essentially no difference between the results of the different protocols presented in Fig.~\ref{fig:app_schedule}. However, from $\Delta/\Omega_0=3$, the behaviors of both $\P$ and $\Q$ start shifting apart. In particular, we observe that the quadratic protocol does not reach a value as high as the linear and cubic protocols do. However, since one should get exactly $\Q=1$ for a QSL state, no protocol reaches the aimed state.

The analysis of the FM-order parameters highlights the similarity between all the results. Indeed, for all protocols considered, $\PFM$ vanishes from $\Delta/\Omega_0 = 3$. Furthermore, even though the exact values of $\QFM$ differ slightly, their  behaviors are qualitatively similar among all protocols. In particular, $\QFM$ vanishes for all considered cases, yet the vanishing point is shifted for the linear protocol. 

While this analysis is not based on global topological metrics such as the topological entanglement entropy, the information brought by the topological operators already gives a valuable indication that small changes in the protocols have no significant impact on the reachable states. We leave a thorough analysis of the optimization of the protocol to maximize the value of $\gamma$ to future work.

\section{\label{sec:App_expressivity} Fidelity analysis of the prepared state }

\begin{figure}[bt]
    \includegraphics[width=\linewidth]{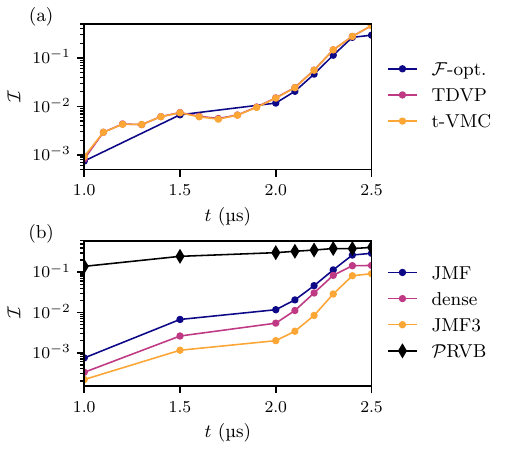}
    \caption{\label{fig:app_fidelity} Infidelity $\mathcal{I}(t) = 1 - |\bra{\psi_\text{exact}(t)}\ket{\psi_\theta(t)}|^2$ of the variational state at all times during the time evolution. 
    (a) Comparison of the $\psi_\theta^\text{JMF}$ Ansatz for various integration schemes, namely fidelity optimization w.r.t. the exact solution (dark blue), exact evolution of the TDVP equation of motion without any sampling (purple) and the usual time-dependent variational Monte Carlo (yellow).
    (b) Comparison of multiple Ansätze trained through fidelity optimization. The Ansätze are the JMF~\eqref{eq:JMF}, the dense Jastrow~\eqref{eq:dense}, the three-body Jastrow~\eqref{app:JMF3} and the projected RVB~\ref{app:RVB}. }
\end{figure}

To better understand the validity of our variational approach, we investigate the main candidates for numerical error in the t-VMC prescription. 
For this purpose, we consider the small lattice of $N=24$ spins on a torus, as depicted in Fig.~\ref{fig:lattice}. 

Firstly, we want to ensure that the chosen numerical scheme, namely t-VMC, converges to the correct state. For this purpose, we compare the fidelity to the exact state at all times for three different numerical schemes: (i) fidelity optimization of our Ansatz with respect to the reference state carried out at every time (no time evolution), (ii) time-dependent variational principle (no Monte Carlo sampling), and (iii) t-VMC as in the main text. We first observe on Fig.~\ref{fig:app_fidelity}(a) that the three considered schemes are qualitatively yielding the same results, with an excellent fidelity at small times, which then increases around $t=\SI{2}{\us}$. 
The discrepancy between TDVP and t-VMC are practically indiscernible, which confirms that our Markov-chain Monte Carlo sampling scheme is not a significant source of numerical error. 
Moreover, the comparison with fidelity optimization allows to rule out the rest of the t-VMC scheme as a potential source of error, in particular the discretization of the dynamical equations and the inversion of the quantum geometric tensor at each step. Thus, calculating and solving the TDVP equation of motion does not increase the infidelity during the simulation. 

Therefore, the only source left to verify is the expressivity of the Ansatz itself. To that aim, we compare Ansätze with varying design and assess whether this has an important impact on the fidelity. We restrict ourselves to Ansätze within the Jastrow class since they allow for a great numerical stability.

The first additional architecture we consider is the usual dense (all-to-all) Jastrow obtained by replacing the invariant parameters of our Ansatz by a dense matrix as $V_{d_{ij}}\to W_{ij}$, leading to
\begin{equation}
    \psi_{\boldsymbol{\theta}}^{\text{dense}}(\boldsymbol{\sigma}) = \exp \left( \sum_{i<j} \sigma_i W_{ij} \sigma_j \right) \prod_i \varphi_i(\sigma_i) \,\text{.}
\label{eq:dense}
\end{equation}

We also consider a more expressive Ansatz by adding a three-body Jastrow interaction term to our existing Ansatz: 
\begin{equation}
    \psi_{\boldsymbol{\theta}}^\text{JMF3}(\boldsymbol{\sigma}) = \exp \left( \sum_{i<j<k} W_{d_{ij}, d_{jk}} \sigma_i \sigma_j \sigma_k \right) \cdot   \psi_{\boldsymbol{\theta}}^\text{JMF}(\boldsymbol{\sigma}) \,\text{.}
\label{app:JMF3}
\end{equation}
Notice that this form is again a translationally invariant Jastrow, supplemented with a mean field. It was shown~\cite{carleo_unitary_2017} that $N$-body Jastrow Ansätze are able to represent states up to a residual involving correlations of order $N>1$. Hence, this Ansatz should perform better than the two-body Ansatz. However, it introduces a significant numerical overhead, as the number of parameters scales quadratically with the system size $N$ instead of linearly. 

Furthermore, another form of variational Ansatz, which can easily represent the RVB state, has been recently proposed in Ref~\cite{giudici_dynamical_2022}. The corresponding wave-function is given by
\begin{equation}
    \ket{\psi_{\boldsymbol{\theta}}^{\mathcal{P}\text{RVB}}} = \bigotimes_i (1+z_2\hat{\sigma}_i^+)(1+z_1\hat{\sigma}_i^-) \ket{\text{RVB}} \,\text{,}
\label{app:RVB}
\end{equation}
where the RVB state is an equal-weight superposition of all defect-free dimerizations of the lattice. The operators are respectively $\hat{\sigma}_i^- = \dyad{g_i}{r_i}$ and $\hat{\sigma}_i^+ = \dyad{r_i}{g_i}$ tuned by the complex parameters $z_1$ and $z_2$. By setting $z_1=0=z_2$, one obtains exactly the sought-after RVB state, which can faithfully be implemented using tensor networks on single vertices and projecting out non-valid dimerizations~\cite{giudici_dynamical_2022}. Whenever $z_1,z_2\neq0$, this Ansatz requires the application of a dense matrix to the RVB state. However, Monte Carlo calculations are efficient only for sparse or local operators (see App.~\ref{sec:App_MCMC}) and thus this Ansatz is not scalable to large systems. 

In order to analyze representational power isolated from any other source of numerical error, these Ansätze are all optimized by minimizing the infidelity to the exact solution $\mathcal{I}(t) = 1 - |\bra{\psi_\text{exact}(t)}\ket{\psi_\theta}|^2$ at all times. The optimizations were done on the small lattice of $N=24$ dispensing from any Monte Carlo sampling. 

We deduce from the results in Fig.~\ref{fig:app_fidelity}(b) that the qualitative behavior between a dense Jastrow Ansatz and an invariant one is the same. While the dense Jastrow has a generally higher fidelity, the JMF Ansatz allows for a great reduction in the number of parameters without significantly impacting the state. This is due to the fact that we include an additional inhomogeneous mean-field part, which breaks any spatial translational invariance.
As expected, the addition of a three-body term increases the expressivity of the Ansatz and reduces even further the infidelity, even though the Ansatz uses invariant parameters (as opposed to the dense Jastrow). For approximately the same number of parameters, the partially invariant three-body Jastrow obtains better results than a dense two-body Ansatz. This reinforces the idea that upon increasing the order the Jastrow, the infidelity of the prepared state can be arbitrarily reduced. 

In contrast, even though the $\mathcal{P}\text{RVB}$ performs comparably to the Jastrow Ansätze at large times, its representativity is significantly worse at intermediate times. Thus, considering t-VMC, the use of this Ansatz would accumulate errors throughout the evolution. Therefore, this Ansatz, though useful in other contexts~\cite{giudici_dynamical_2022}, is unsuitable for the simulation of dynamical preparation protocols from trivial initial states. 

All considered Ansätze can represent exactly the RVB state. For the $\mathcal{P}$RVB Ansatz, this is achieved for $z_1=0=z_2$, while for the two-body Jastrow this is obtained in the limit $W\to-\infty$ (see Sec.~\ref{sec:App_ansatz}). Still, none reaches vanishing infidelities at the final times of the simulation of the state preparation. Therefore, we conclude the final states are different from the RVB state. Thus, when designing an expressive Ansatz for this task, it is not only key for it to be able to represent the RVB state, but also to be able to capture the departure from it induced by the long-range tails in the van der Waals potential which ultimately produces a non-trivial topologically ordered phase at times $t>\SI{2.0}{\us}$.
In general, we conclude that our choice of 2-body translation invariant Jastrow with inhomogeneous mean field yields a sufficiently high representational power to represent all states encountered during the time evolution.

\section{\label{sec:App_noise} Stochastic unraveling of disordered open quantum systems }
\begin{figure*}
    \includegraphics[width=\linewidth]{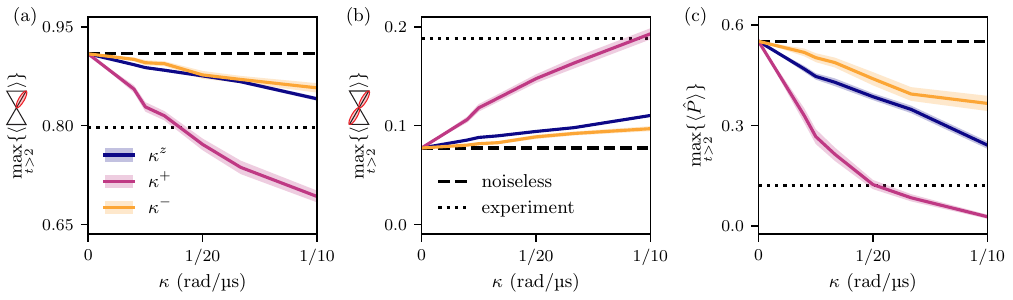}
    \caption{\label{fig:app_channels} Effect of each noise channel, as defined in Eq.~\eqref{eq:channels}, as a function of its angular rate. Each color corresponds to the simulation of the protocol subject to one noise channel at a time and in the absence of spatial disorder.
    The lighter bands indicate the root mean square error over the quantum trajectories. All plotted quantities correspond to the maximal values inside the QSL-like phase for $t>\SI{2.0}{\us}$. Noisy simulations are carried out on the smaller lattice ($N=24$) presented in Fig.~\ref{fig:lattice}. The results are compared to the noiseless exact evolution on the same lattice (dashed) and to the experimental results from Ref.~\cite{semeghini_probing_2021} obtained on the real-scale lattice in Fig.~\ref{fig:lattice} (dotted).
    (a) Maximal probability to find dimers on the vertices of the lattice. (b) Maximal probability to find double dimers (defects). (c) Maximal value of the diagonal operator $\P$. 
    }
\end{figure*}

The t-VMC prescription solves the time-dependent Schrödinger equation of a closed quantum system. However, upon considering the interactions with the environment, the latter needs to be treated within the formalism of open quantum system~\cite{breuer_theory_2007}. This can be done through a Lindblad master equation. Such an exact description implies the time evolution of the density matrix $\hat{\rho}$ under a Liouvillian map. The complexity of this evolution scales quadratically worse than that of $\ket{\psi}$, making the approach intractable for most use cases, in particular in the present one which involves over $200$ Rydberg atoms. To circumvent this, we make use of the quantum trajectories prescription, which expresses the solution of the master equation $\hat{\rho}$ as an average over multiple random realizations (trajectories) $\ket{\psi}$ of a stochastic Schrödinger equation. 

\subsection{\label{sec:App_algo} Algorithm}
To efficiently compute trajectories, we use the faster-than-the-clock algorithm~\cite{johansson_qutip_2012, daley_quantum_2014} described in what follows. The state of each trajectory is evolved under the pseudo-Hermitian Hamiltonian 
\begin{equation}
    \hat{\mathcal{H}}^{\text{nh}} = \hat{\mathcal{H}} - \frac{i}{2} \sum_j \hat{L}_j^\dagger \hat{L}_j \,\text{,}
\end{equation}
where the sum runs over all noise channels $j$ with associated jump operators $\hat L_j$. Due to the non-Hermitian nature of this Hamiltonian, the norm of the wave function will decrease in time. We make use of this norm to decide when a quantum jump occurs. For each trajectory, a random number $\eta\in[0,1]$ is drawn and once the norm reaches $||\psi(t)||^2=1-\eta$, a quantum jump occurs. The jump is randomly picked from all the existing channels with respective probabilities $p_j(t) = \ev{\smash{\hat{L}_j^\dagger \hat{L}_j}}/\eta$. The state after a jump is given by
\begin{equation}
    \ket{\psi(t^+)} = \frac{1}{\sqrt{\eta p_j(t^-)}}\hat{L}_j\ket{\psi(t^-)} \,\text{,}
\end{equation}
which results in a normalized wave function. After the state is projected, a new number $\eta$ is drawn, corresponding to the trigger for the next jump. 

In order to evaluate operators on a mixed state, no explicit reconstruction of the density matrix is needed. Instead, one can average the expectation value of the observables of interest over as many stochastic trajectories as needed to reach convergence. In our case, we use $100$ trajectories, each for an independent random realization of the spatial disorder in the parameters of the physical system (see next section for details on the modeling of disorder). This procedure optimally reduces the total variance on the trajectory and spatial-disorder average~\cite{vicentini_optimal_2019}.

\subsection{Sources of error}

We consider two sources of error: spatial disorder on the parameters of the Hamiltonian and noise induced by the coupling to the environment.

We start by considering inhomogeneities on the Hamiltonian, to represent the disorder in the experiment. This is accounted for as an imprecision in the local fields applied on each atom: $\hat{\mathcal{H}}(t)\to\hat{\mathcal{H}}(t) - \frac{\Omega(t)}{2}\sum_i \omega_i \hat{\sigma}^x_i -\Delta(t) \sum_i \delta_i \hat{n}_i$, where each shift $\omega_i,\delta_i$ is randomly picked from a normal distribution with chosen variance $\mathcal{N}(0,\sigma_{(n,x)}^2)$~\cite{ebadi_quantum_2021}. Since we are considering the errors on the fields to be normally distributed, the impact on observables is reduced as compared to that of systematic errors~\cite{cai_stochastic_2023}, which we do not consider here. As $\Omega(t)$ is only associated with the kinetic part of the Hamiltonian, inhomogeneities have a small effect on the resulting state. In contrast, modifying the driving field $\Delta(t)$ induces a decoherence effect~\cite{kropf_effective_2016}. 

Furthermore, we consider the principal error channels, whose jump operators are defined as follows:
\begin{equation}
\begin{aligned}
    \hat{L}^{(z)}_i &= \sqrt{\kappa^{z}} \left(\dyad{r_i}{r_i} - \dyad{g_i}{g_i}\right) \,\text{,} \\
    \hat{L}^{(+)}_i &= \sqrt{\kappa^{+}} \dyad{r_i}{g_i} \,\text{,} \\
    \hat{L}^{(-)}_i &= \sqrt{\kappa^{-}} \dyad{g_i}{r_i} \,\text{,}
\label{eq:channels}
\end{aligned}
\end{equation}
with respective error rates $\kappa^j$. 

To understand the real impact of each noise channel on the prepared state, we conduct a qualitative analysis presented in Fig.~\ref{fig:app_channels}. Therein, we chose to solely consider diagonal observables in order to avoid the noise effects induced by the basis rotation in the experimental results, which are hard to predict. 
Firstly, we observe that all noise channels have a similar effect on the quantities shown in Fig.~\ref{fig:app_channels}, namely bringing the results of simulations into closer line with those of experiments, as expected. 
We observe that out of the three error channels considered, scattering has the strongest impact on the observables. Indeed, as the density of excitations does not exceed $\ev{\hat n} \approx 1/3$ during the whole time evolution, the application of scattering strongly modifies the mean occupation, and thus, the observables depending on it. By generating excitations randomly in the state, it creates new dimers which do not respect any kind of Rydberg blockade and thus increases the density of double dimers and imperfect dimerizations. 
On the other hand, as the decay channel destroys excitations, it generates imperfect dimerizations but does not create states violating the Rydberg constraint. Thus, it does not greatly influence the density of defects, but impacts $\P$ slightly. 
Moreover, decoherence has the most difficult effect to predict. As these observables are all diagonal, the influence is reduced compared to $\Q$, yet it still contributes to a shift from the exact noiseless results. 

Finally, we compare these effects to the experimental measurements obtained in Ref.~\cite{semeghini_probing_2021}. We find a match between our noisy results and the experimental ones for a value of the scattering rate in agreement with that measured on the experimental setup ($\kappa^+_\text{exp} = 1/(\SI{150}{\us}) \approx \SI{1}{\radian}/\SI{23.9}{\us}$). Notice however that the results of Fig.~\ref{fig:app_channels} only involve isolated noise channels on a small lattice of $N=24$ atoms, not the $N=219$ lattice of the actual experiment.

In Sec.~\ref{sec:noise} of the main text, we confirm the predictive character of our modeling of the errors in the experimental protocol upon using the error rates of Ref.~\cite{semeghini_probing_2021, ebadi_quantum_2021}. For this purpose, we consider scattering processes at an angular rate of $\kappa^+= 2\pi/(\SI{150}{\us})$ and decay at $\kappa^-=2\pi/(\SI{80}{\us})$. Furthermore, as previously introduced, we use inhomogeneities in the fields to be of $\sigma_x^2 = 3\%$ for $\Omega(t)$ and $\sigma_n^2 = 2\%$ for $\Delta(t)$. The results obtained for this setting are the ones presented and discussed in Sec.~\ref{sec:noise}. 

\section{\label{sec:App_TEE} Lattice topology and finite size effects}
\begin{figure}
    \includegraphics[width=\linewidth]{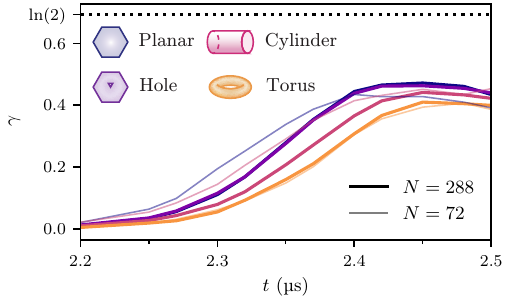}
    \caption{\label{fig:app_topology} Topological entanglement entropy obtained throughout the time evolution for various geometries of the lattice: the main lattice with open boundary conditions as presented in Fig.~\ref{fig:lattice} (Planar), the same lattice with a hole in the middle (Hole), a cylindrical (Cylinder) and a toric one (Torus), with respectively one and two periodic boundaries. The TEE is calculated for similar partitions as depicted in Fig.~\ref{fig:lattice}, optimally centered in the bulk of each lattice. Solid lines indicate simulations conducted on a system of $N=288$ particles, as the lighter lines are for smaller systems of $N=72$. The planar lattice with a hole was not simulated for a lower number of particles as such a lattice would not possess any bulk. 
    }
\end{figure}

As discussed in Sec.~\ref{sec:entropy}, when studying and describing topological states, the geometry and topology of the lattice (size, boundary conditions, genus) are of prime importance. The great flexibility of our method, unlike previous state-of-the-art techniques, enables us to directly investigate the dependence of the topological features of the system, namely the topological entanglement entropy, on its geometry and topology. 
To avoid finite-size effects, we consider larger lattices than in the main text with $N=288$ atoms. Doing so, we are able to consider a lattice with open boundary conditions (similar to the one presented in Fig.~\ref{fig:lattice}) and a hole in the middle acting as an effective boundary, while the system still possesses a bulk to estimate the TEE. We firstly observe on Fig.~\ref{fig:app_topology} that while the system is bigger than that of the main text and topological inequivalent, one observes a comparable value for the TEE $\gamma = 0.463(3)$, not closer to $\ln(2)$. 

In order to conclude on finite size effects, we additionally consider a smaller lattice of $N=72$, with same boundary conditions. We observe a qualitatively similar behavior for the three topologically identical planar lattices. Despite proving the presence of topological order through the finite value of TEE, the value of $\gamma$ for the smaller system is generally diminished, only attaining $\gamma = 0.435(4)$ at its peak. As topological order is rooted in long-range entanglement (see Sec.\ref{sec:entropy}), this behavior highlights the susceptibility of $\gamma$ to finite size effects. 
However, as the consideration of a larger lattice than in the main text does not improve the topological entanglement entropy, reaching only $\gamma=0.472(3)$, we are able to rule out finite-size effects as the root for departure from the characteristic $\gamma$ value of \Ztwo order. 

For a perfect QSL, the boundary conditions and the genus of the lattice only affect the degeneracy of the ground state of the system, having no impact on the value of the TEE. We verify this in Fig.~\ref{fig:app_topology}, where no qualitative difference is found between the topological entanglement entropy of the two lattices with same boundary conditions but different genus ($0$ for the planar lattice and $1$ for the one with a hole). However, lattices with periodic boundary conditions reach slightly lower $\gamma$ values, whereas systems defined on a cylinder possess higher topological order than the ones on a torus.

A potential explanation for this reduced value of $\gamma$ could lie in the long-range tails of the Rydberg potential. Indeed, these strongly affect the ground state of the system (see Sec.~\ref{sec:state_prep}) and still have significant effects on the outcome of dynamical preparation, as analyzed by Ref.~\cite{sahay_quantum_2023}, though mitigated upon a proper choice of protocol. 
Thus, the presence of periodic boundary conditions allows for additional long-range interactions across periodic images, thereby emphasizing this destabilizing effect on a torus as compared to lattices with open boundary conditions, ultimately yielding a lower value of $\gamma$.

\bibliography{main}

\end{document}